\documentclass[fleqn,usenatbib]{mnras}

\usepackage{newtxtext,newtxmath}

\usepackage[T1]{fontenc}

\DeclareRobustCommand{\VAN}[3]{#2}
\let\VANthebibliography\thebibliography
\def\thebibliography{\DeclareRobustCommand{\VAN}[3]{##3}\VANthebibliography}

\usepackage{graphicx}	
\usepackage{amsmath}	
\usepackage[normalem]{ulem} 

\usepackage{pdflscape} 

\title[BTFR in MaNGA and IllustrisTNG]{A Comparison of the Baryonic Tully-Fisher Relation in MaNGA and IllustrisTNG}

\author[J. S. Goddy et al.]{
Julian S. Goddy,$^{1,2}$\thanks{E-mail: jg3837@drexel.edu (JSG)}
David V. Stark,$^{1,3}$
Karen L. Masters,$^{1}$
Kevin Bundy,$^{4}$
Niv Drory,$^{5}$
David R. Law$^{3}$
\\
$^{1}$Departments of Physics and Astronomy, Haverford College, 370 Lancaster Avenue, Haverford, PA 19041, USA \\
$^{2}$Department of Mechanical Engineering and Mechanics, Drexel University, 3141 Chestnut Street Philadelphia, PA 19104 USA \\
$^{3}$Space Telescope Science Institute, 3700 San Martin Dr., Baltimore, MD 21218, USA \\
$^{4}$UCO/Lick Observatory, University of California, Santa Cruz, 1156 High St. Santa Cruz, CA 95064, USA\\
$^{5}$McDonald Observatory, The University of Texas at Austin, 1 University Station, Austin, TX 78712, USA\\
}

\date{Accepted XXX. Received YYY; in original form ZZZ}

\pubyear{2021}

\begin{document}
\label{firstpage}
\pagerange{\pageref{firstpage}--\pageref{lastpage}}
\maketitle

\begin{abstract}
We compare an observed Baryonic Tully-Fisher Relation (BTFR) from the Mapping Nearby Galaxies at Apache Point Observatory (MaNGA) and HI-MaNGA surveys to a simulated BTFR from the cosmological magnetohydrodynamical simulation IllustrisTNG. To do so, we calibrate the BTFR  of the local universe using 377 galaxies from the MaNGA and HI-MaNGA surveys, and perform mock 21 cm observations of matching galaxies from IllustrisTNG.  The mock observations are used to ensure that the comparison with the observed galaxies is fair since it has identical measurement algorithms, observational limitations, biases and uncertainties. For comparison, we also calculate the BTFR for the simulation without mock observations, and demonstrate how mock observations are necessary to fairly and consistently compare between observational and theoretical data. We report a MaNGA BTFR of log$_{10} (M_{ \rm Bary}/M_\odot)= (2.97 \pm 0.18)$  log$_{10} V_{ \rm Rot} + (4.04 \pm 0.41)\,\log_{10}{M_{\odot}}$ and an IllustrisTNG BTFR of log$_{10} (M_{ \rm Bary}/M_\odot) =  (2.94 \pm 0.23$) log$_{10} V_{ \rm Rot} + (4.15 \pm 0.44)\,\log_{10}{M_{\odot}}$. 
Thus, MaNGA and IllustrisTNG produce BTFRs that agree within uncertainties, demonstrating that IllustrisTNG has created a galaxy population that obeys the observed relationship between mass and rotation velocity in the observed universe. 
\end{abstract}




\section{Introduction}

\label{sec:Intro}

The Baryonic Tully-Fisher Relation (BTFR), i.e. the strong correlation between total galaxy \textit{baryonic} mass (stars and cold gas) and rotation speed \citep{Mcgaugh2000}, has proven to be one of the most fundamental scaling relations for disk galaxies. Unlike its relative predecessor, the Tully-Fisher Relation (TFR), that relates \textit{luminosity} and rotation speed \citep{TullyFisher1977}, the BTFR appears to follow a single power law over several orders of magnitude in baryonic mass, from dwarf to giant galaxy scales.

The BTFR/TFR have found many uses within the field of extragalactic astronomy. An immediate application was as a distance indicator, for which it is still used at present day \citep[e.g.,][]{Tully2012}. Additionally, the BTFR has been employed to test how the baryonic/dark matter mass fraction differs in barred vs. unbarred galaxies \citep{Courteau2003}, to estimate the total baryonic mass within galaxies \citep{McGaugh2015,Wang2015, Sorce2015}, to constrain the stellar Initial Mass Function (IMF; \citealt{Stark2009}), to study formation scenarios for specific sub-classes of galaxies that deviate from the main relation \citep[][]{Lelli2015a,Ogle2019} and to investigate how the line-of-sight velocities of Mg II absorbers relate to the rotation velocity of stellar disks \citep{Diamond-Stanic2016}. 

Given the tightness of the BTFR and its reliability over several orders of magnitude in galaxy mass, it has regularly been employed as a test of cosmological and galaxy formation models. Models using $\Lambda$ Cold Dark Matter ($\Lambda$CDM) cosmological frameworks predict a slope that depends on the assumed values of free parameters and definitions but is consistently between 3 and 4 \citep{Bradford2016, Lelli2019}. In contrast, Modified Newtonian dynamics \citep{Milgrom1983} predicts a BTFR slope of exactly 4.0 with no scatter \citep{Bradford2016, Ogle2019}.  

Complex hydrodynamical cosmological models employ a variety of prescriptions for star formation, turbulence, and energetic feedback, often that are implemented below their nominal resolution \citep{Schaye2015}. Comparing simulated galaxy populations to real ones is the only means of testing whether their prescriptions are reliable. Historically, hydrodynamical simulations have been able to qualitatively reproduce a BTFR-like power law between baryonic mass and rotation speed, but until very recently matching the true relationship {\it quantitatively} has been challenging \citep{Somerville2015}. One possible recent success is the reported agreement between the BTFR calibrated with the SPARC observed galaxy sample and the SIMBA hydrodynamical galaxy formation simulation \citep{Glowacki2020}. While this result used ``ideal" rotation curves from the simulation rather than mock observations, which may complicate the comparison, the followup work \citep{Glowacki2021} did present a BTFR based on mock observations which appeared to show good agreement. Other examples of success in the last five years include the complimentary works \citet{Ferrero2017,Sales2017} who explored the relation in APOSTLE/EAGLE simulations.

While different galaxy formation models can generate different BTFRs, the observational determination of the ``true" BTFR is also subject to a variety of systematic errors. For instance, there is considerable observational scatter, which \citet{Andersen2003} determine is related to kinematic asymmetries in particular galaxies. Moreover, competing methods about how to map luminosity to stellar mass (e.g., choice of IMF) result in systematic uncertainties of at least a factor of two in the BTFR scaling. Similarly, there are differing methods and uncertainties when estimating the characteristic rotation velocities of galaxies: HI linewidths --- for which there is no standard measurement algorithm \citep{Springob2005} and which may be biased in very gas poor galaxies since the gas disk does may not adequately fill the potential well --- and rotation curves where authors debate over whether to use the maximum velocity versus the flat velocity \citep[e.g][]{Catinella2005, Stark2009,Ponomareva2018,Lelli2019}. Furthermore, some rotation curves fail to reach any clear maximum or flat part, and all velocity measurements are subject to uncertainties in inclination corrections (especially for low-mass galaxies; \citealt{Masters2006, Bradford2016}). \citet{Lelli2016} find that even in simulations, linewidths and rotation curves give notably different BTFRs. A calibrated BTFR is also sensitive to the details of the sample used and the linear fitting algorithm \citep{Meyer2008, Bradford2016}. 

Due to these multiple potential causes for variations in a derived BTFR, the best way to compare BTFRs is via direct statistical comparison between data sets with commensurable properties and measurement techniques, rather than simply comparing BTFR fits \citep{Bradford2016}. Following this philosophy, we investigate whether the IllustrisTNG \citep{Weinberger2017,Naiman2018, Nelson2017, Marinacci2018, Pillepich2018a, Pillepich2018b, Springel2018} cosmological hydrodynamical simulation accurately reproduces the observed BTFR based on observational data from the SDSS-IV Sloan Digital Sky Survey (SDSS; \citealt{Blanton2017}), Mapping Nearby Galaxies at Apache-Point Observatory (MaNGA; \citealt{Bundy2015, Drory2015, Law2015, Yan2016a, Yan2016b, Law2016, Wake2017, Law2021}) and the HI-MaNGA \citep{Masters2019, Stark2021} surveys. To account for observational and sample selection biases, we apply mock single-dish 21cm observations to the IllustrisTNG data set and select the sample to match the mass and star formation properties of the main observational sample. This approach not only ensures a fair and consistent comparison between these observed and simulated datasets, but also tests the necessity of the mock observations for comparing these two datasets. 

This work organized as follows: In Section~\ref{sec:methods}, we describe the data, derived parameters, and linear fitting algorithm used to calibrate the BTFR using MaNGA and IllustrisTNG data. In Section~\ref{sec:results}, we present the derived BTFRs, and in Section~\ref{sec:discussion}, we compare the resulting BTFR calibrations to each other and to those from the literature, and discuss the impact of conducting mock observations on simulation data. Our conclusions and potential future directions are summarized in Section~\ref{sec:conclusions}. Throughout this work we report all uncertainties to 1-$\sigma$. Where needed to generate physical quantities we use the default cosmological parameters used by the MaNGA Pipe3D data products ($H_0=71 \text{ km s}^{-1}\text{Mpc}^{-1}; \Omega_m=0.27; \Omega_\Lambda=0.73.$) and IllustrisTNG ($H_0=67.7 \text{ km s}^{-1}\text{Mpc}^{-1}; \Omega_m=0.31; \Omega_\Lambda=0.69.$). This slight difference in $H_0$ for our low redshift samples will lead to differences of order $h^2$ in values, or $<10$\%. This offset may impact the zero-point of the BTFR, but not the slope.

\section{Methods}
\label{sec:methods}
\subsection{The BTFR in MaNGA}
\subsubsection{The Observational Data Set}
\label{section:MaNGAdetails}

We calibrate the BTFR using observational data from the Mapping Nearby Galaxies at Apache-Point Observatory (MaNGA; \citealt{Bundy2015, Drory2015, Law2015, Yan2016a, Yan2016b, Law2016, Wake2017, Law2021}) and the HI-MaNGA surveys \citep{Masters2019, Stark2021}. The combination of these two surveys is ideal for this study; they provide a large sample of galaxies that are close enough that we can accurately determine their properties---including stellar mass, HI-mass, and rotation velocity. 

MaNGA is a large spectroscopic survey of 10,000 nearby galaxies (z $<$ 0.06) using the BOSS spectrograph \citep{Smee2013} on the 2.5m Apache Point Optical Telescope (APO; \citealt{Gunn2006}), and is part of the 4th generation Sloan Digital Sky Survey (SDSS; \citealt{Blanton2017}). Specifically we use the 10th MaNGA Product Launch (MPL-10; \citealt{Law2021})

We use the MaNGA ``Data Reduction Pipeline" (DRP; \citealt{Law2016}), which contains the calibrated and reduced data. MaNGA collaborators have also created ``value-added-catalogs" (VACs) that build on the primary SDSS photometry and spectroscopy. HI-MaNGA, a 21cm follow-up campaign of all $z<0.05$ MaNGA galaxies using the Green Bank Radio Telescope (GBT) and existing ALFALFA survey data \citep{Haynes2018}, is one of these VACs. We make use of data in the second release of HI-MaNGA \citep[DR2; ][]{Stark2021}. It contains analyzed HI gas properties such as integrated HI masses, linewidths, and recession velocities. Pipe3D \citep{Sanchez2016b, Sanchez2018} is another VAC that we use. It is an analysis pipeline that employs spatial and spectral binning and analysis on the MaNGA datacubes to extract analyzed stellar and ionized gas properties.

\subsubsection{Derived Quantities}
\label{section:derivedQuantities} 

In this section, we describe our derived parameters and selection criteria for the MaNGA dataset.  

Our total baryonic mass is given by:  
\begin{equation}
 M_{\rm Bary} = M_{\rm HI} + M_{\rm He} + M_{\rm H_2} + M_{\rm \star} = \frac{4}{3} \: M_{\rm HI} +1.07 \: M_\star 
 \label{eq:linearTotalMass}
\end{equation}
where $M_{\rm Bary}$ is the total baryonic mass, $M_{\rm HI}$ is the total neutral atomic Hydrogen (HI) gas mass, $M_{\rm He}$ is the total Helium gas mass, $M_{\rm H_2}$ is the total molecular Hydrogen gas mass (H$_2$), and  $M_\star$ is the total stellar mass.

HI mass ($M_{\rm HI}$) is calculated as in \citet{Stark2021}:
\begin{equation}
 \frac{M_{\rm HI}}{M_\odot} = \frac{2.36 \times 10^5}{(1+z)^2}\bigg(\frac{D}{\text{Mpc}}\bigg)^2 \bigg(\frac{F_{\rm HI}}{\text{Jy km s}^{-1}}\bigg)
 \label{eq:linearHIMass}
\end{equation}
where $z$ is the redshift of the galaxy, $D$ is the luminosity distance in Mpc, and $F_{\rm HI}$ is the integrated flux of the HI line in Jy km s$^{-1}$ from the HI-MaNGA survey. We multiply our HI masses by 4/3 to account for the contribution from Helium since the universe is roughly 75\% Hydrogen (\citealt{McGaugh2012}; this factor is not reflected in Eq~\ref{eq:linearHIMass}). We do not include any correction for HI self-absorption, but this would only have a minor impact on HI masses (the corrections never exceed 20\%).

The observational uncertainty on the HI mass ($\sigma_{M_{\rm HI}}$) is calculated directly from the uncertainty on the HI flux:
\begin{equation}
 \sigma_{M_{\rm HI}} = M_{\rm HI} \frac{\sigma_{F_{\rm HI}}}{F_{\rm HI}},
\label{eq:linearHImassuncertainty}
\end{equation} 
where $\sigma_{F_{\rm HI}}$ is the HI flux uncertainty and $F_{\rm HI}$ is the total HI flux. 

Stellar masses are calculated in Pipe-3D using a composite stellar population, which is fit to each spaxel spectrum taking into account the contribution of dust extinction \citep{Sanchez2016a,Sanchez2016b}. These values are derived from spectroscopy, which gives a good handle on the star formation history (SFH). These values are limited to within the MaNGA integral field unit (IFU), but are in good agreement with the global stellar masses from DRP estimated via broadband photometry\footnote{\url{https://www.sdss.org/dr17/manga/manga-data/manga-pipe3d-value-added-catalog/}}. Accordingly, even though the DRP photometric estimates are global and so lack aperture effects, the choice of DRP or Pipe3D stellar masses will not have a significant impact on the analysis in this paper.  

Pipe3D provides estimates of the stellar mass uncertainty but they are formal statistical uncertainties which arise from observational uncertainties and do not account for large systematic errors like the choice of IMF, which is assumed to be Salpeter \citep{Salpeter1955}. Determining these systematic uncertainties on the stellar masses would involve making mock stellar spectra and running them through stellar population fitting routines, which is beyond the scope of this project. Accordingly, we add 0.2 dex to the reported statistical integrated stellar mass uncertainties, $\sigma_{M_\star}$, so that our final, corrected stellar mass uncertainty, $\sigma_{M_{\star,c}}$, is given as
\begin{equation}
\sigma_{M_{\star,c}} = \sqrt{(\sigma_{M_\star})^2+(0.2)^2} 
\label{eq:linearStellarMassuncertainty} 
\end{equation}
This additional systematic uncertainty accounts for factors such as the specific stellar population model used and the choice of IMF and is consistent with variations between different stellar mass estimation algorithms \citep{Kannappan2007}.

H$_2$ is difficult to measure directly due to its stability and symmetry \citep{Bolatto2013} but is traditionally traced using emission from carbon monoxide (CO), and then converted using a scaling relation between CO flux and H$_2$ mass. However, this relation has considerable uncertainty \citep{Bolatto2013} and we lack CO observations for the majority of our data set. \citet{McGaugh2020} estimate H$_2$ content by combining two scaling relations --- one converting from stellar mass to star formation rate, and another converting from star formation rate to H$_2$ mass --- finding that H$_2$ mass in a galaxy is well approximated by 7\% of the stellar mass. However, there can be substantial variation in the actual H$_2$ content of galaxies, even at the same stellar mass \citep{Catinella2018}, so we stress that the 7\% correction is only true on average. Although H$_2$ is usually negligible compared with the stars and HI in a galaxy in terms of mass, it is becoming increasingly important to quantify the amount of H$_2$ as the accuracy of the other measurements, particularly the stellar mass, improve \citep{McGaugh2020}. Accordingly, we apply this simplistic estimate and multiply our stellar masses by 1.07 to account for the contribution of H$_2$. This method was chosen over the relations between SFR and $H_2$ mass in \citet{Leroy2013} since while it would be straightforward for the MaNGA data, it would be complicated to implement for the IllustrisTNG data. Therefore, it would not be easy to have a consistent method across both samples.

The uncertainty on the baryonic mass is then given by
\begin{equation}
  \sigma_{M_{\rm Bary}} = \sqrt{\bigg[\frac{4}{3} \big(\sigma_{M_{\rm HI}} \big) \bigg]^2 + \bigg[1.07 \big(\sigma_{M_{\star,c}} \big) \bigg]^2} 
 \label{eq:linearMassuncertainty}
\end{equation}

We use HI linewidths to estimate rotation velocities, specifically the HI linewidth measured at 50\% of the peak using a linear fit for both sides of the HI profile ($W_{\rm F50}$). Some authors prefer to use linewidth measured at 20\% of the peak, so we calculate a $V_{\rm Rot, 20}$ similarly, just using $W_{\rm F20}$ (see Section~\ref{sec:comparison} for further discussion). We correct linewidths for inclination, cosmological broadening, and turbulent motions using 
\begin{equation}
 V_{\rm Rot}= \bigg[\frac{W_{\rm F50}-2\Delta v \lambda}{1+z}-\Delta t\bigg]\frac{1}{2\text{ sin} \: i}
 \label{eq:Wrot}
\end{equation}
where $V_{\rm Rot}$ is the corrected rotation velocity, and $W_{\rm F50}$ is the uncorrected linewidth, $\Delta v = 5.00$ km s$^{-1}$ is the effective resolution before Hanning smoothing. $\lambda$ accounts for the impact of noise on the effective resolution in simulations by \citet{Springob2005} and is given by 
\begin{equation}
 \lambda = \begin{cases}
 0.005 & \text{log(SNR}) < 0.6 \\
 -0.4685+0.785\;\text{log(SNR})&0.6 \leq\text{log(SNR})<1.1\\ 
 0.395 & \text{log(SNR}) \geq 1.1 
 \end{cases}
 \label{eq:lambda}
\end{equation}
where SNR is the signal to noise ratio calculated as ($f_{\rm peak}-rms)/rms$, where $f_{\rm peak}$ and $rms$ are the peak flux density and root mean squared noise level from the HI-MaNGA catalog respectively. $z$ is the optical galaxy redshift. $\Delta t = 6.5$ km s$^{-1}$ is a correction for turbulent motions (also from \citealt{Springob2005}). We use $i$ for the inclination -- the angle of the galaxy to our line of sight and is calculated using: 
\begin{equation}
 \text{cos} \: i = \sqrt{\frac{(b/a)^2-q^2}{1-q^2}}
 \label{eq:inclination}
\end{equation}
where $q=0.2$ is a reasonable estimate for disc thickness \citep{Masters2014} and $b/a$ is the observed axial ratio of the galaxy. We use the elliptical Petrosian axial ratio taken from the NASA Sloan Atlas (NSA\footnote{\url{https://www.sdss.org/dr17/manga/manga-target-selection/nsa/}}).
This value for $V_{\rm Rot}$ is 1/2 of that given by equation 4 of \citet{Masters2019}. We have divided the expression by two in order to estimate the rotation velocity as opposed to a linewidth. 

The uncertainty on $V_{\rm Rot}$, $\sigma_{\rm V}$, is twice the uncertainty on the centroid of the HI line detection, $\sigma_{V_{\rm HI}}$. The uncertainty on $i$, $\sigma_i$ is a 30\% fractional uncertainty. This is an educated guess of its typical uncertainty that 
is combination of errors in the assumption about q, and in measuring the observed axial ratio (b/a) driven by galaxy structure.

We assume that the uncertainties on $\Delta v, \lambda, z, \text{ and } \Delta t$ are negligible compared to the uncertainties on $W_{\rm F50}$ and $i$. The uncertainty on the rotation velocity is then given by

\begin{equation}
\sigma_{\rm V_{Rot}} = \sqrt{\bigg[\frac{\partial V_{\rm Rot}}{\partial W_{\rm F50}} \big(\sigma_{\rm V} \big) \bigg]^2 + \bigg[\frac{\partial V_{\rm Rot}}{\partial i} \big(\sigma_i \big) \bigg]^2 }
\label{eq:linearLinewidthuncertainty}
\end{equation}

where from Equation \ref{eq:Wrot}, 
\begin{gather*}
 \frac{\partial V_{\rm Rot}}{\partial W_{\rm F50} }= \frac{1}{2(1+z)\text{ sin} \: i}\\
 \frac{\partial V_{\rm Rot}}{\partial i} = \bigg[\frac{W_{\rm F50}-2 \Delta v \lambda}{1+z}-\Delta t\bigg] \frac{-1}{2\text{ tan} \: i \text{ sin} \: i} \\
\end{gather*}

\subsubsection{Sample Selection}
\label{section:sampleSelection}

In order to create a representative and reliable sample of galaxies to which we can fit the BTFR, we need to apply some sample cuts. We start with the cross match of the HI-MaNGA DR2 and Pipe-3D analysis of MPL-10. This contains 3635 galaxies, but roughly half of the HI-MaNGA galaxies are non-detections and lack HI mass and linewidth estimates. We then make the following sub-selections, and note how many galaxies remain assuming we make the cuts in the following order:

\begin{itemize}
   \item We ensure that galaxies have measured values for all of the parameters in Section \ref{section:derivedQuantities}. 1710 galaxies satisfy this criteria.
   
    \item We select galaxies with inclination $>$ 30$\degr$ in order to minimize the effect of inclination uncertainties (which have 1/sin($i$) dependence). This leaves 1442 galaxies. 
    
    \item We select galaxies with a ratio of HI mass to stellar mass $>$ 0.3 so that there is enough HI to fill the potential well and extend to the flat part of each galaxy's rotation curve. 714 galaxies satisfy this criteria. 
    
    \item We select galaxies with 21cm line detection signal to noise (``SNR" in HI-MaNGA) $>$ 5 so that we can accurately pick out the signal from the noise and to avoid extremely large linewidth uncertainties. This leaves 540 galaxies. 
    
    \item We remove HI confused galaxies. HI confusion occurs when multiple galaxies are close enough in position and velocity that they blend into a single 21 cm source and cause these measurements to be unreliable. We do this by removing galaxies that are flagged as being confused (by ``conflag" in HI-MaNGA). 
\end{itemize}
  This selection results in a final sample of 377 galaxies.  

The mass ratio cut was done in order to systematically remove strong outliers by visual inspection before trying to fit a line to the data. This was motivated by the significant scatter in the data (see Section \ref{sec:MCMC}). We also performed tests on the MaNGA data and found that this cut does not significantly change the BTFR fit, but does slightly reduce scatter and consequently it facilitates the linear fit of the data. Although this cut introduces a strong selection bias towards low-mass galaxies, we note that the simulated IllustrisTNG sample is equivalently biased since it is created from galaxies that match the observed MaNGA sample (see Section \ref{section:BTFRinTNG}).  

\subsection{The BTFR in IllustrisTNG}
\subsubsection{The Simulation}
\label{section:TNGdetails}

The cosmological magnetohydrodynamical simulation IllustrisTNG \citep{Weinberger2017,Naiman2018, Nelson2017, Marinacci2018, Pillepich2018b, Pillepich2018a, Springel2018} is a recent state-of-the-art cosmological  simulation. It self-consistently models the evolution of gas and dark matter from the early universe until today. It does so via the moving-mesh mode \texttt{AREPO} \citep{AREPO2010}, which numerically solves the coupled system of partial differential equations describing gravitational and electromagnetic fields, radiative cooling of gas, the conversion of gas into stars, the subsequent evolution of those stars, the chemical enrichment of the interstellar, circumgalactic, and intergalactic media, and the formation, growth, and energetic feedback of supermassive black holes in both low and high accretion states \citep{Pillepich2018a}. It is important to note that IllustrisTNG simulates a completely contiguous volume in the universe, not just isolated galaxies.

To test the success of IllustrisTNG in recreating the BTFR, we use the API version of IllustrisTNG100 (called ``TNG100-1"), which simulates the evolution of a periodic cube with side lengths 110.7 Mpc and baryonic mass resolution $1.4 \times 10^6 \: M_\odot$ and dark matter resolution $7.5 \times 10^6 \: M_\odot$. We use snapshot 99, corresponding to $z=0$, which matches the approximate redshift of our observational sample.

\subsubsection{Calibrating the BTFR using IllustrisTNG}
\label{section:BTFRinTNG}

We calibrate the BTFR using a sample of IllustrisTNG galaxies with matching properties to our observed MaNGA sample. Specifically, we consider an IllustrisTNG galaxy a match to a MaNGA galaxy if both galaxies have consistent stellar masses and star formation rates within their estimated observational uncertainties. For details on how IllustrisTNG calculates stellar masses and star formation rates, see \citealt{Pillepich2018a}).
The observed star formation rate is the integrated star formation rate derived from the integrated H$\alpha$ flux in Pipe3D. The observed stellar masses are described in Section \ref{section:MaNGAdetails}. We match star formation rates as a simple check that the galaxies have similar amounts of gas. If multiple IllustrisTNG galaxies satisfy the above matching criteria for a particular MaNGA galaxy, we pick the IllustrisTNG galaxy with the closest stellar mass to the MaNGA galaxy.

\subsubsection{Mock Observations}
\label{sec:MockObservationMethods}

Since the goal is to compare the BTFR in IllustrisTNG to the one from MaNGA, we must carefully consider the impact of the observational limitations of MaNGA and HI-MaNGA observations which are not present within simulations. Specifically, due to the apparent faintness of galaxies, it is necessary to use large telescopes to maximize the photon collecting area and therefore create a detectable signal. Even still, all real observations have finite detection limits. Furthermore, single-dish radio telescopes observing at 21cm have primary beams that are substantially larger (e.g., $9\arcmin$ for the GBT) than the typical MaNGA galaxy ($\sim$30\arcsec), meaning all gas information is unresolved.  

In contrast, there are no observational limitations on simulations. Accordingly, IllustrisTNG can create detailed maps of the gas in each galaxy, which it uses to calculate, for example, the gas mass and rotation speed of the galaxy. 

These two datasets, the unresolved gas measurements and detailed gas maps provide different types of information. Comparing them is not a fair comparison between similar quantities.

Since we do not have access to detailed HI maps for our observed galaxies, we have to create mock unresolved 21cm spectra for IllustrisTNG galaxies, allowing for a fair comparison between real and mock galaxies. Mock observations are performed using the \texttt{Martini} package\footnote{\url{https://github.com/kyleaoman/martini}; version 1.0.1} \citep{Martini2019}. \texttt{Martini} calculates the HI mass fraction for each gas particle, modeling self-shielding from the metagalactic ionizing background following \citet{Rahmati2013}, and correcting for an empirical pressure-dependent molecular gas fraction according to \citet{Blitz2006}. HI emission from each gas particle is modeled with a Gaussian line profile centered on the particle velocity and assuming that it is optically thin (so each observed particle has flux proportional to its mass).

After the HI component of each simulated galaxy is modeled, it is ``observed" by assuming that the galaxy is at the distance of its observational counterpart. We first create a data cube with 1 arcminute spatial resolution and 10 ${\rm km\,s^{-1}}$ velocity resolution for each galaxy, and then convolve it with the on-sky sensitivity profile of the GBT, assumed to follow a Gaussian with a FWHM of $9\arcmin$ and a peak value of unity. This creates an integrated HI spectral profile for each simulated galaxy. We then add random Gaussian noise with a standard deviation equivalent to the rms noise of the observational counterpart.

This process allows us to pass the simulated spectra into the \textit{same} data analysis algorithm used to measure properties (HI mass, linewidth) of the real HI-MaNGA observations, and we include identical corrections for contributions from He and H$_2$, as well as line broadening. We run these data-reduced flux and rotation velocity measurements through the same Bayesian MCMC linear fitting algorithm that we used to obtain the observational MaNGA BTFR described in Section \ref{sec:MCMC}

Thus, the gas properties of our simulated data set are measured {\it in the exact same way} as the real observational sample, with an attempt to reproduce identical observational biases and measurement algorithms. However, we note that we do not perform mock observations on the stellar masses as that would require performing mock spectrophotometry of IllustrisTNG galaxies, which is beyond the scope of this project. Accordingly, we assume that if we had measured the stellar masses in the IllustrisTNG galaxies with MaNGA they would have the same uncertainties as the corresponding MaNGA galaxies since they are set to be at the same distances and have matching stellar masses and star formation rates. 

In order to illustrate the effect and importance of mock observations, we also calibrate the IllustrisTNG BTFR using the rotational velocity measured with the complete 3D spatial/velocity information provided by IllustrisTNG (i.e without the mock observations). Specifically, we use ``SubhaloVmax", which we call $V_{\rm Max}$, defined as the ``maximum value of the spherically-averaged rotation curve."\footnote{\url{https://www.tng-project.org/data/docs/specifications/}}. $V_{\rm Max}$ is known ``perfectly" within the simulation, not subject to observational biases, making it a useful foil to the BTFR calibrated using mock HI linewidth measurements. 

\begin{figure*}
\centering
\includegraphics[angle=0,scale=0.35]{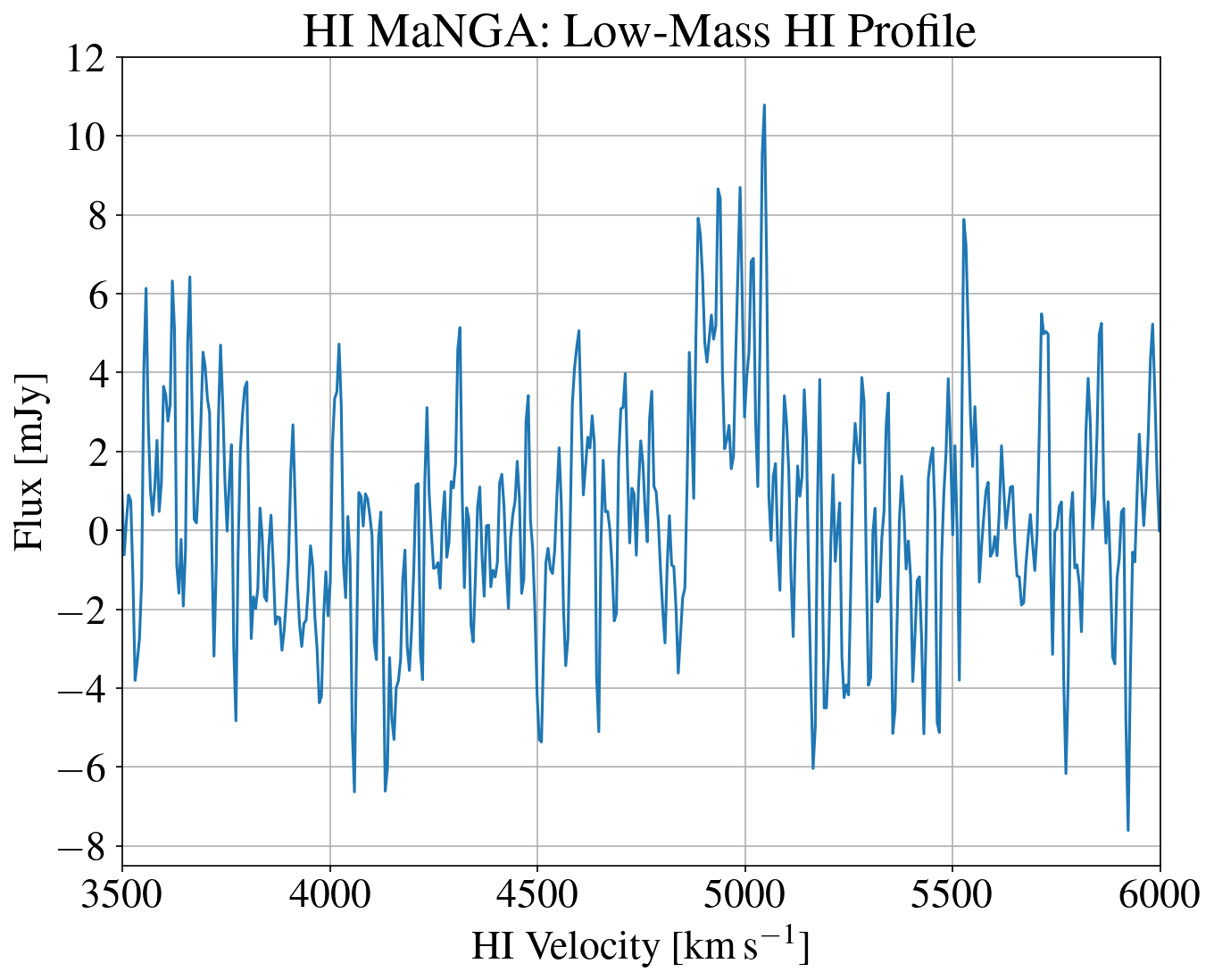}
\includegraphics[angle=0,scale=0.35]{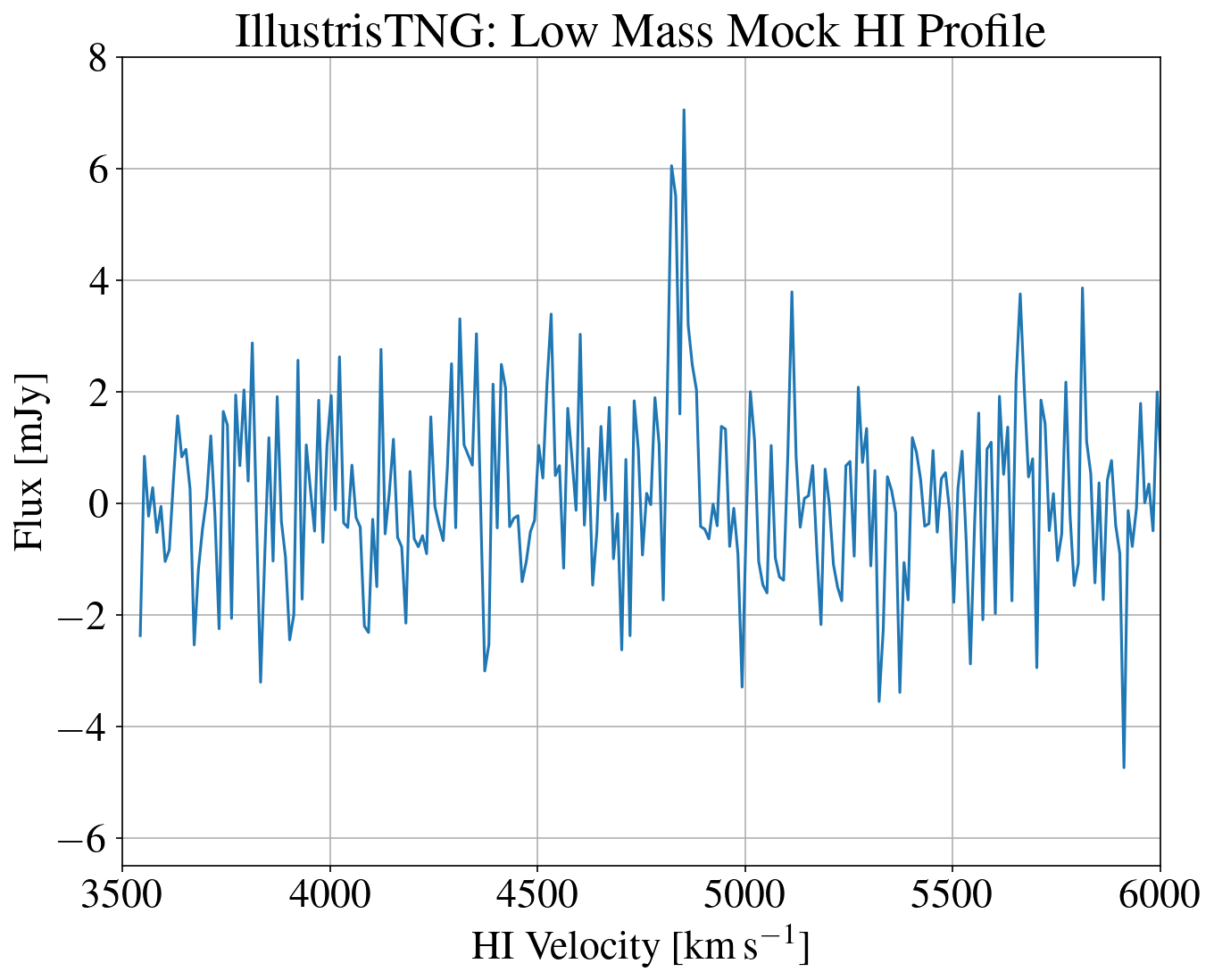}
\includegraphics[angle=0,scale=0.35]{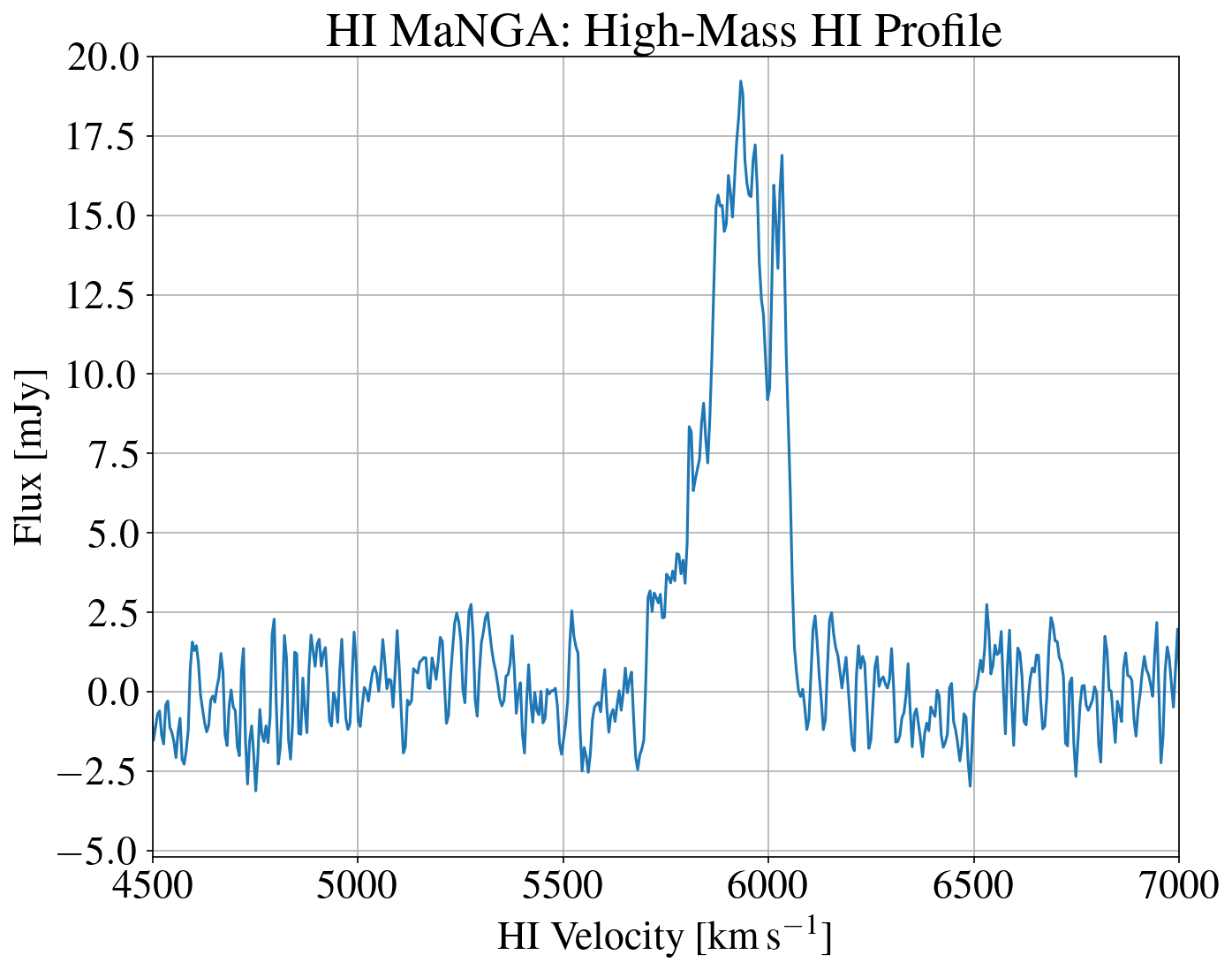}
\includegraphics[angle=0,scale=0.35]{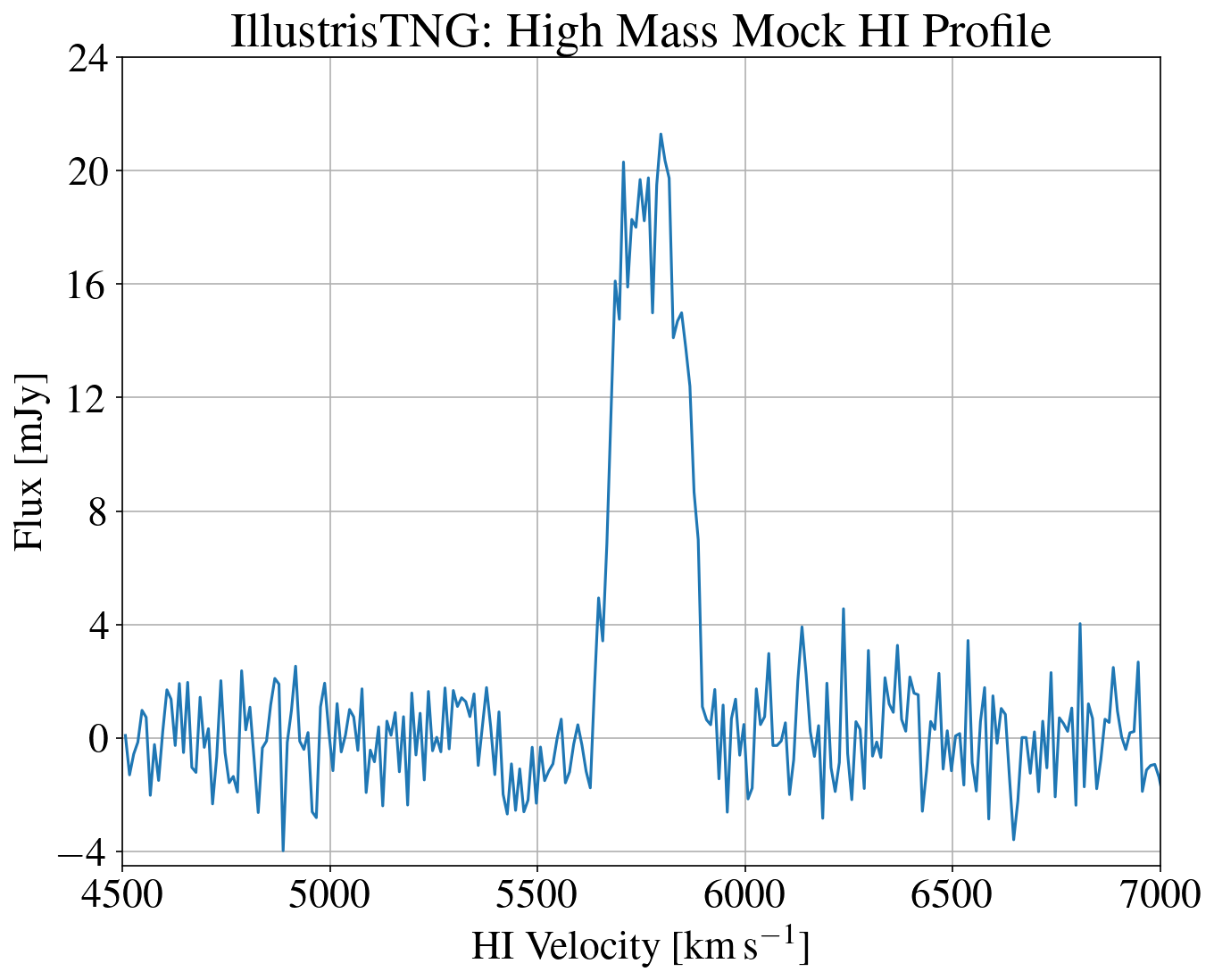}

\caption{ Sample matching HI profiles for low mass galaxies (MaNGA plateifu: 8937-6102 on top left and IllustrisTNG ID: 583726 on top right) and high mass galaxies (MaNGA plateifu: 8548-3702 on bottom left and IllustrisTNG ID: 586848 on bottom right)}
\label{figure:HI-profile-comparison}
\end{figure*}

\begin{figure}
\centering
\includegraphics[angle=0,scale=0.3]{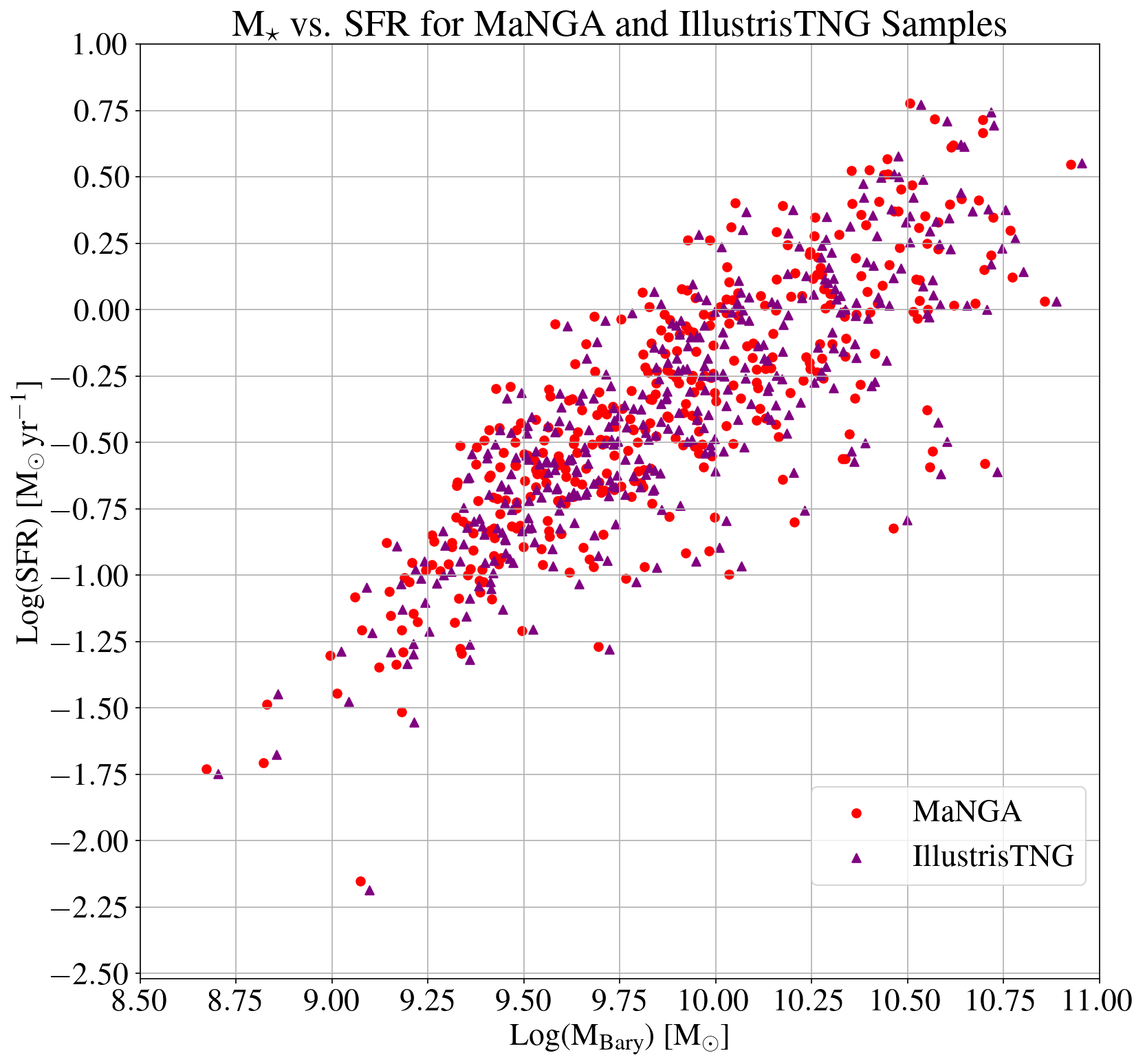}

\caption{Stellar Mass vs. Star Formation Rate for the MaNGA and IllustrisTNG samples showing the similarity of the primary and control samples.}
\label{figure:M_star-SFR-comparison}
\end{figure}

The purpose of the mock observations is to introduce all the observational biases from the HI-MaNGA data into the IllustrisTNG simulated data so that the comparison between the two datasets is fair and consistent. We show sample HI profiles for matching galaxies (according to the matching criteria in Section \ref{section:BTFRinTNG}) from HI-MaNGA and IllustrisTNG in Figure \ref{figure:HI-profile-comparison} and plot the stellar mass vs. star formation rate in Figure \ref{figure:M_star-SFR-comparison} to demonstrate their similarity.

\subsection{The Bayesian MCMC Linear Fitting Algorithm}
\label{sec:MCMC}

A complication in calibrating the BTFR with the MaNGA data is fitting a line to data with significant observational scatter and/or strong outliers. This scatter is most likely a combination of intrinsic scatter and observational uncertainty. A $\Lambda$CDM cosmological framework predicts significant intrinsic scatter in the BTFR \citep{Bradford2016}, although the true level of intrinsic scatter is debated \citep{McGaugh2012,Lelli2016,Papastergis2016}. The observational scatter could be caused by random measurement uncertainty as well as systematic effects, e.g., HI poorly tracing the rotation velocity due to it not filling the potential well of gas-poor galaxies \citep{Lelli2015b} or other systematic variations in the relationship between HI linewidth and the ``flat" part of the rotation curve \citep{Verheijen2001}. 

To address potential complications when fitting a line to data with considerable scatter, we use a Bayesian Markov Chain Monte Carlo (MCMC) linear regression fitting algorithm
based on the \texttt{theano} and \texttt{PYMC3} modules in \texttt{Python} \citep{Theano2016, Salvatier2015}. This algorithm identifies likely significant outliers and removes them from the data set prior to performing a linear fit to the remaining data. For for more details see \citet{Hogg2010}; we use the implementation by Coleman Krawcyzk\footnote{\url{https://github.com/CKrawczyk/jupyter\_data\_languages/blob/master/mcmc\_fit\_with\_outliers\_pymc.ipynb}}. 

We add an additional variance prior and a corresponding standard deviation to account for the intrinsic scatter caused by intrinsic differences between individual galaxies. Accordingly, the model has five free parameters: slope, y-intercept, intrinsic scatter on outlier and inlier points, and measurement uncertainty. We use normal distributions as priors for the slope and y-intercept and inverse gamma functions for the intrinsic variance on the outlier and inlier points. We also change the target percentage that random walks will be added to the chain (called \texttt{target\_accept}) from its default value of 0.8 to 0.999. This ensures that the algorithm only selects points along its random walk, which helps it converge. 

The MCMC algorithm used in this work conducts fits which determine best fit linear parameters of $y$ vs. $x$ by minimizing the scatter in the $y$ direction. However, as there is no clear {\it dependent} variable in the BTFR, we prefer to treat the two variables symmetrically. Therefore, we use the MCMC algorithm to fit both rotation velocity vs. baryonic mass (forward fit) and baryonic mass vs. rotation velocity (inverse fit). We then combine these results into a final best-fit relation (bisector fit) using the bisector method described in \citet{Isobe1990}.

\section{Results} 
\label{sec:results}

We now present the derived Baryonic Tully-Fisher relations for the observational MaNGA sample as well as the mock-observed (and simulated) IllustrisTNG samples. We show the forward, inverse, and bisector fits in all cases. Our analysis here always uses $V_{50}$ for the velocity variable. In Section~\ref{sec:comparison}, we will discuss how alternative velocity measurements can affect our results. To provide a comparison between relations derived from observational versus purely theoretical quantities, in Section ~\ref{sec:need_mocks} we also perform the same fitting routines on IllustrisTNG data but using $V_{\rm max}$, a rotation velocity measure derived directly from the simulations and not subject to measurement uncertainty.  All derived best fit parameters and their associated uncertainties are provided in Table~\ref{tab:BTFRwithAndWithoutTurbulence}. 

\subsection{The MaNGA BTFR} 

\begin{figure*}
\centering
\includegraphics[angle=0,scale=0.3]{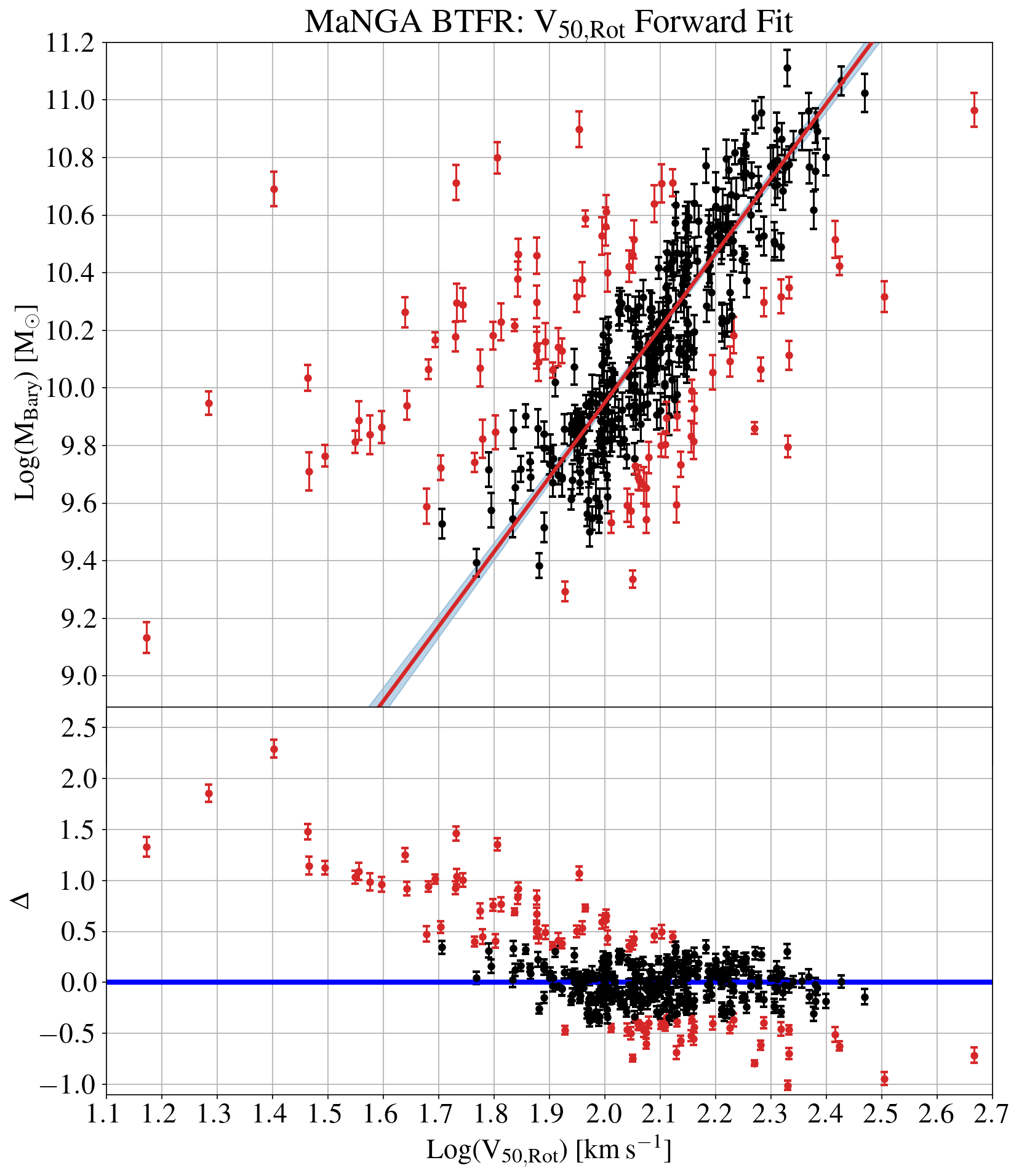}
\includegraphics[angle=0,scale=0.3]{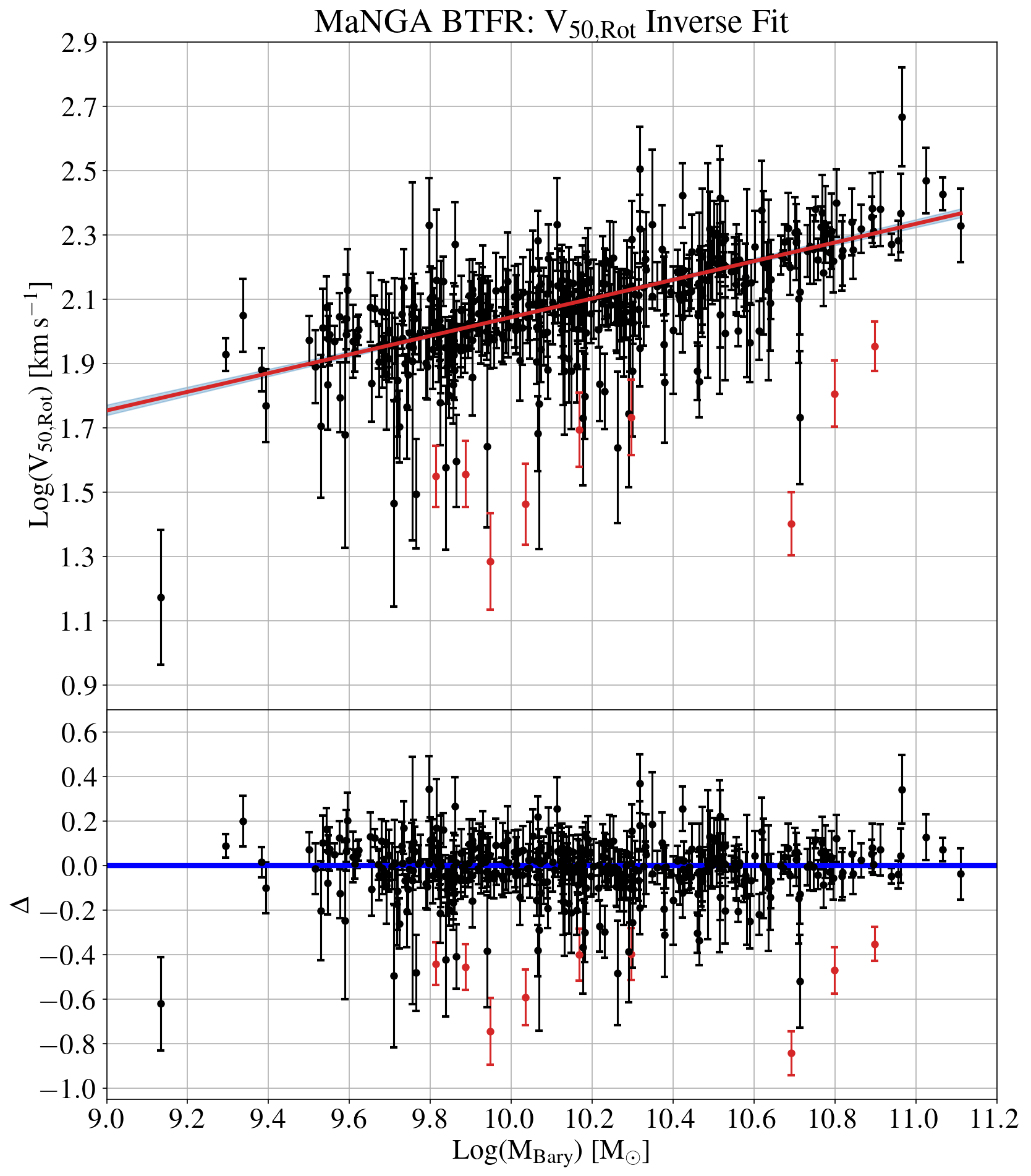}
\includegraphics[angle=0,scale=0.3]{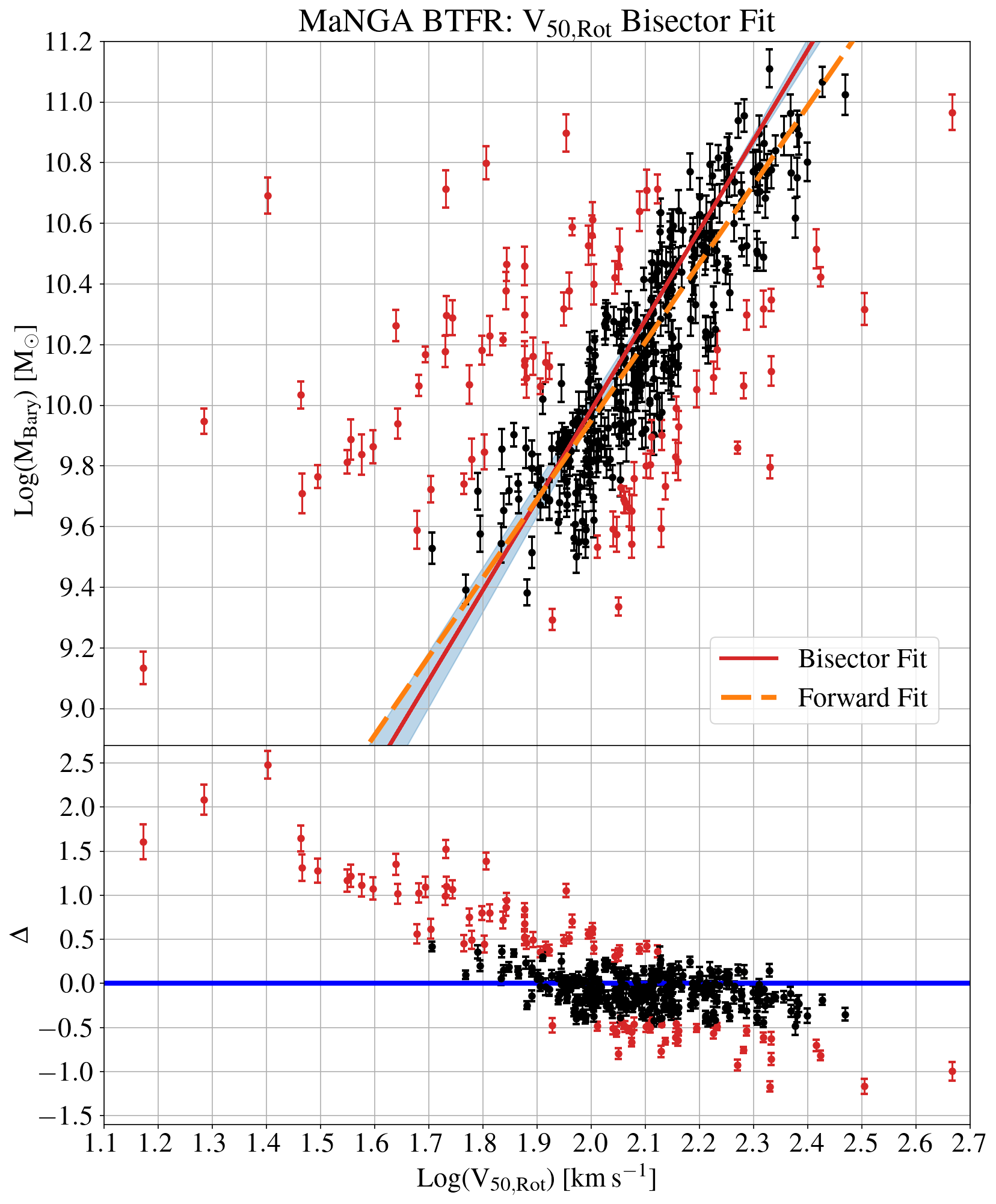}

\caption{The MaNGA BTFR forward fit with rotation velocity on the horizontal axis and baryonic mass on the vertical axis (top left), the inverse fit with baryonic mass on the horizontal axis and rotation velocity on the vertical axis (top right), and the bisector fit with rotation velocity on the horizontal axis and baryonic mass on the vertical axis (bottom). The points that the Bayesian MCMC linear regression code determined have $>80\%$ probability of being an outlier are plotted in red. We also show the 1-$\sigma$ best fit region, and the best fit lines in blue, and red respectively. The subplots show the residuals for each point using the same color criteria as for the main plot and a horizontal line at zero for reference. The forward and bisector fits have 88 outliers and the inverse fit has 9 outliers. The best fit lines, clockwise from left, have slope 2.59 $\pm$ 0.09 with a y-intercept of 4.76 $\pm$ 0.19, slope of 3.44 $\pm$ 0.16 with a y-intercept of 2.95 $\pm$ 0.33, and 2.97 $\pm$ 0.18 with a y-intercept of 4.04 $\pm$ 0.41.}
\label{figure:MaNGA_BTFR}
\end{figure*}

We show the rotation velocity vs. baryonic mass (top left) baryonic mass vs. rotation velocity (top right) and the bisector fit (bottom) with rotation velocity on the horizontal axis and baryonic mass on the vertical axis for the MaNGA sample in Figure~\ref{figure:MaNGA_BTFR}. The points that the Bayesian MCMC linear regression code determined have $>80\%$ probability of being outliers are plotted in red. We also show the 1-$\sigma$ best fit region, and the best fit lines in blue, and red respectively. The subplots show the residuals for each point using the same color criteria as for the main plot and a horizontal line at zero for reference. 

The resulting bisector fit has a slope of 2.97 $\pm$ 0.18 with a y-intercept of $\log(M_{\rm bary}/M_\odot) = 4.04 \pm$ 0.41.

\subsection{The IllustrisTNG BTFR}
\label{sec:TNGBTFRdetails}

\begin{figure*}
\centering
\includegraphics[angle=0,scale=0.3]{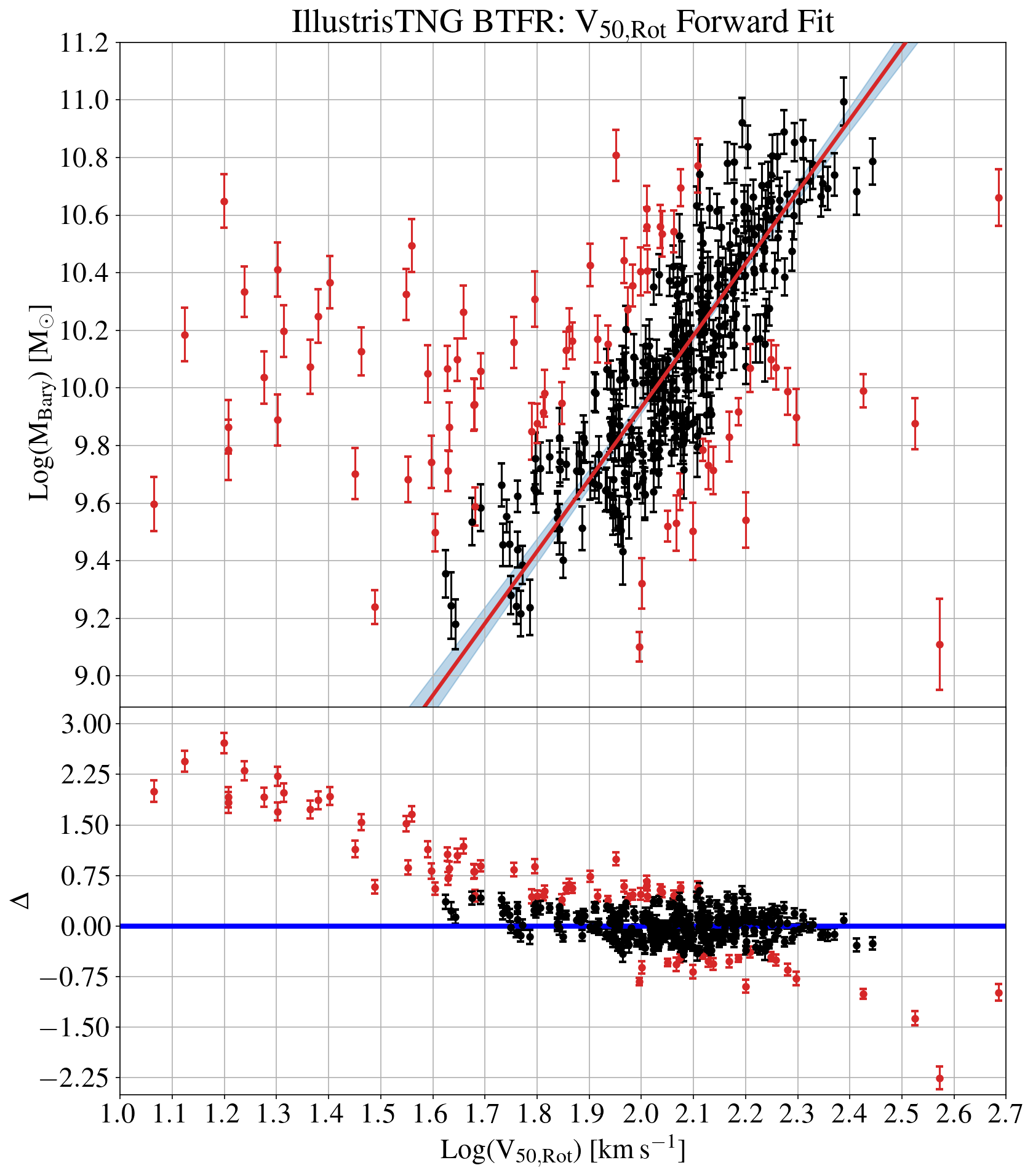}
\includegraphics[angle=0,scale=0.3]{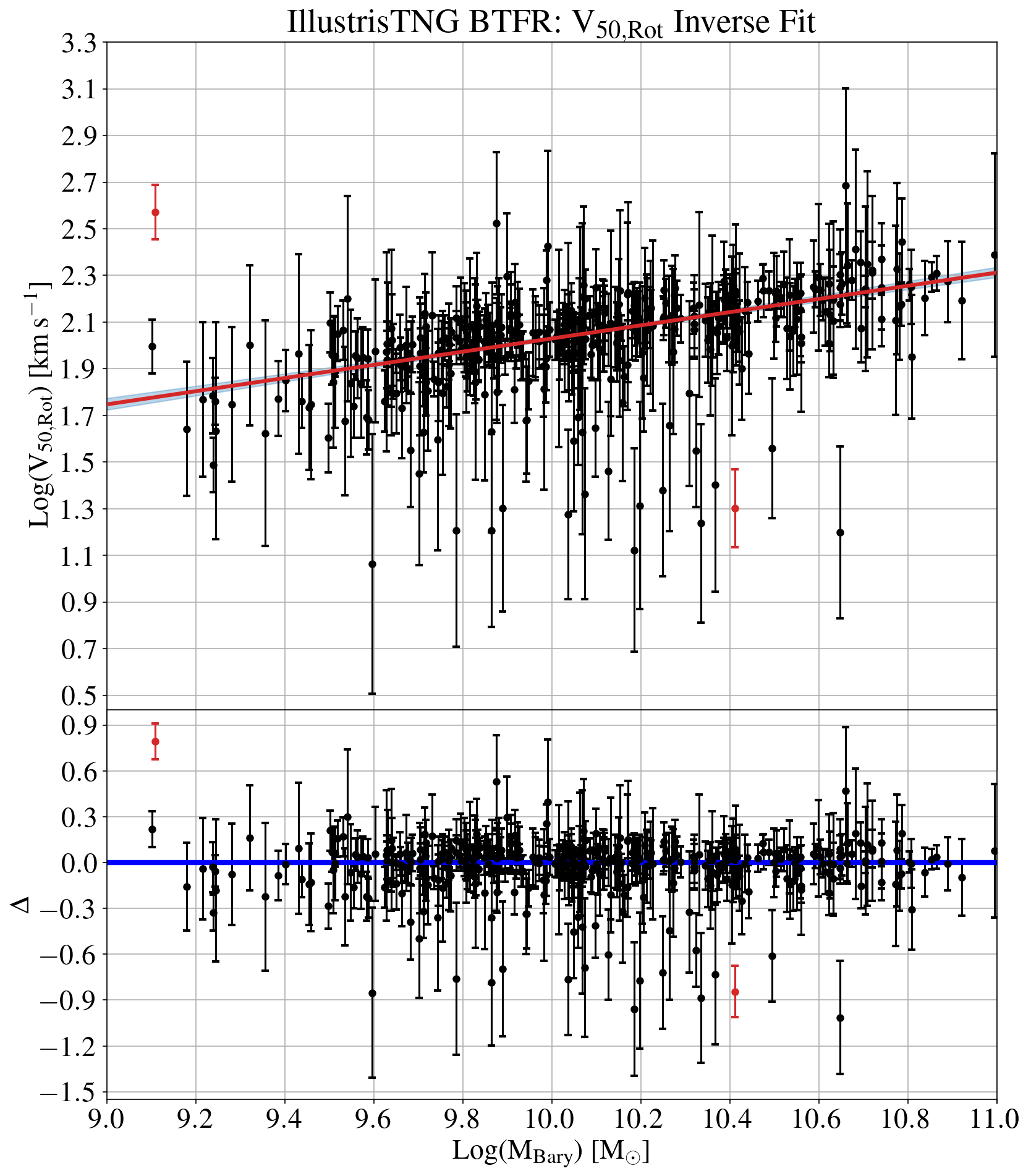}
\includegraphics[angle=0,scale=0.3]{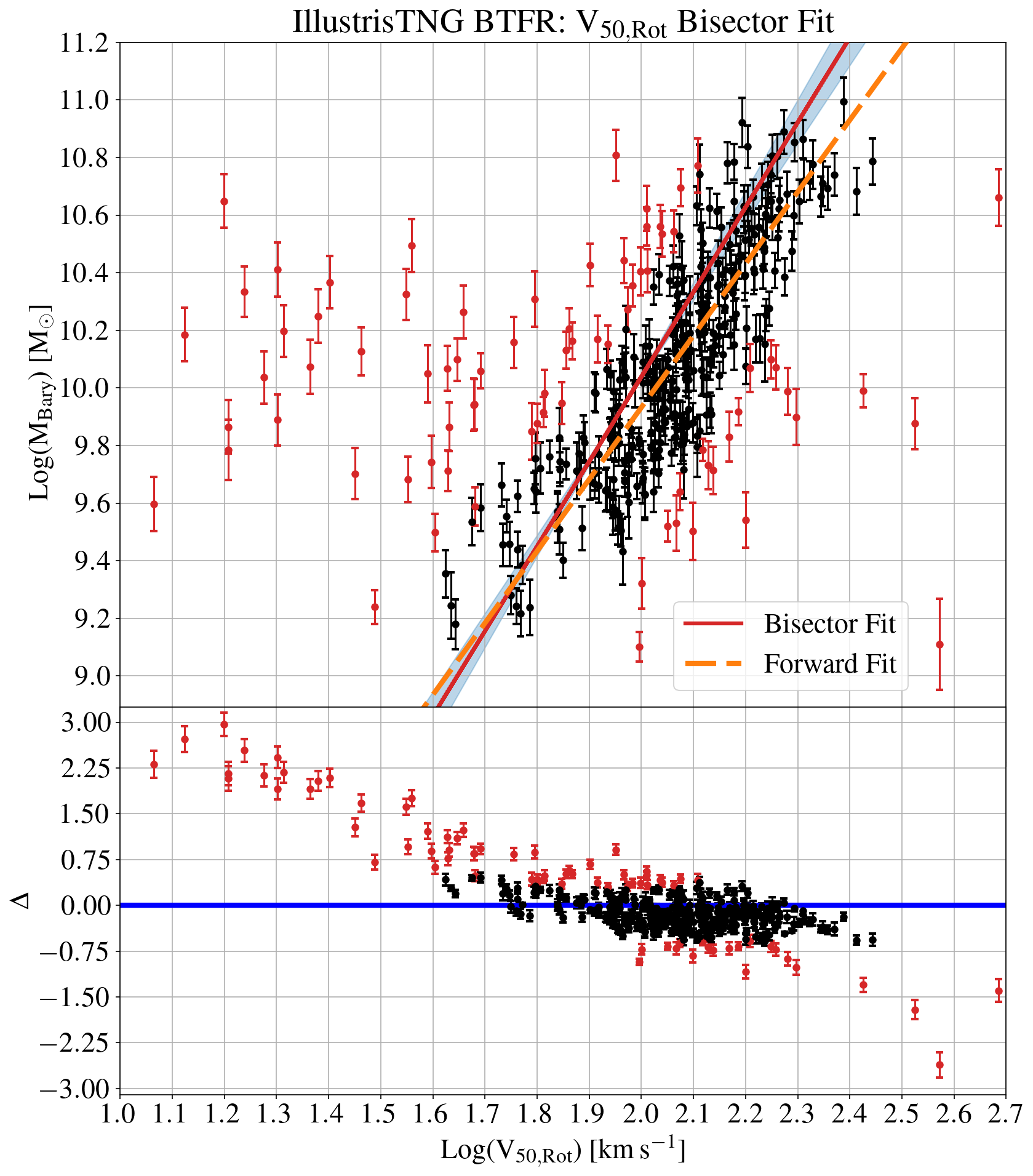}
\caption{Same as Figure~\ref{figure:MaNGA_BTFR} but for the IllustrisTNG mock sample. The forward and bisector fits have 78 outliers and the inverse fit has 2 outliers. The best fit lines, clockwise from left, have slope 2.50 $\pm$ 0.13 with a y-intercept of 4.94 $\pm$ 0.28, slope of 3.56 $\pm$ 0.28 with a y-intercept of 2.78 $\pm$ 0.57, and 2.94 $\pm$ 0.23 with a y-intercept of 4.15 $\pm$ 0.44.}
\label{figure:TNG_BTFR}
\end{figure*}

Figure~\ref{figure:TNG_BTFR} illustrates the mock observational data and resulting BTFR fit using IllustrisTNG. The plots show the same information as Figure~\ref{figure:MaNGA_BTFR}, but for the IllustrisTNG mock-observed sample instead of the MaNGA sample. 
The resulting bisector fit has a slope of 2.94 $\pm$ 0.23 with a y-intercept of $\log(M_{\rm bary}/M_\odot) = 4.15 \pm$ 0.44.

\section{Discussion}

\label{sec:discussion}

\begin{figure*}
\centering
\includegraphics[angle=0,scale=0.3]{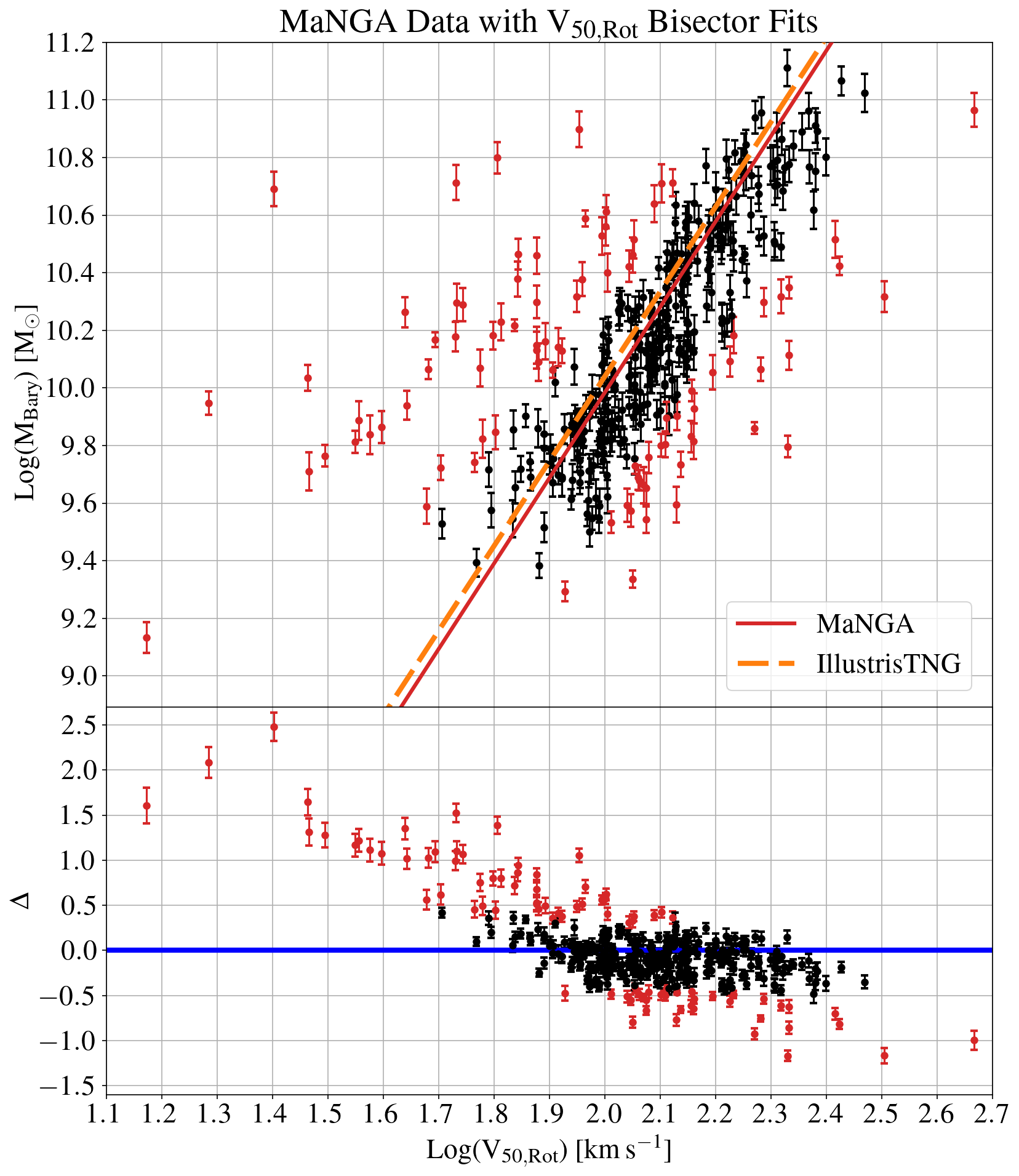}
\includegraphics[angle=0,scale=0.3]{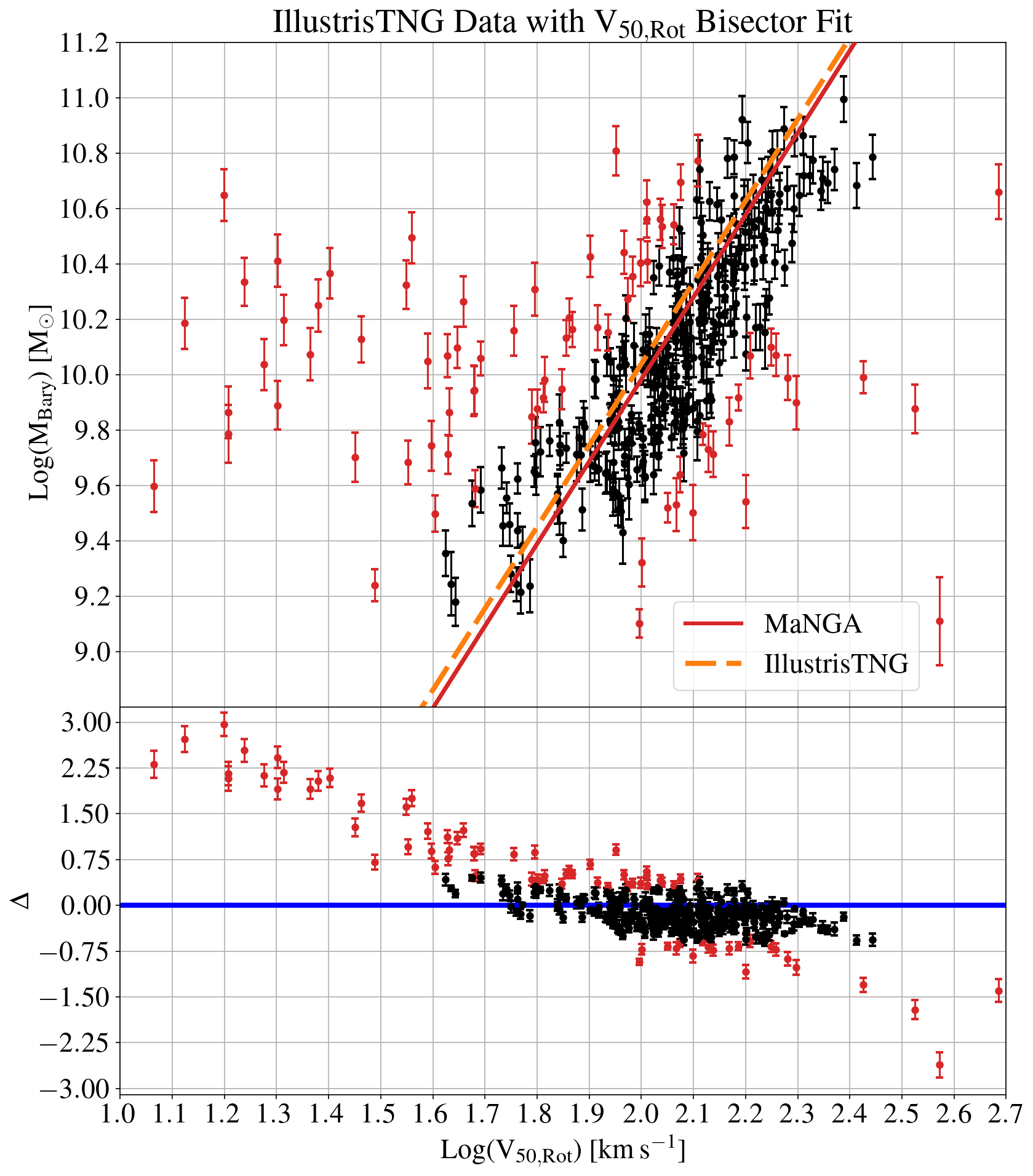}

\caption{The MaNGA data (left) and IllustrisTNG (right) calculated using V$_{\rm 50,Rot}$ with both the MaNGA (solid) and IllustrisTNG (dashed) V$_{\rm 50,Rot}$ BTFR bisector fits overplotted. We use the same color criteria as for figures \ref{figure:MaNGA_BTFR} and \ref{figure:TNG_BTFR}, except we do not show the respective 1-$\sigma$ regions in these plots for clarity. Rotation velocity is on the horizontal axis and baryonic mass is on the vertical axis. The MaNGA BTFR bisector fit has a slope of 2.97 $\pm$ 0.18 with a y-intercept of 4.04 $\pm$ 0.41 and the IllustrisTNG BTFR bisector fit has a slope of 2.94 $\pm$ 0.23 with a y-intercept of 4.15 $\pm$ 0.44.}
\label{figure:MaNGA_TNG_bisector}
\end{figure*}

MaNGA and IllustrisTNG produce BTFRs that agree within uncertainties, lending support to the idea that IllustrisTNG has created a galaxy population that obeys the observed relationship between mass and rotation velocity in the observed universe. We demonstrate this agreement by plotting both the MaNGA and IllustrisTNG BTFR bisector fits on the MaNGA data and the IllustrisTNG data, which are shown in Figure \ref{figure:MaNGA_TNG_bisector}.

 While we have made every effort to make a fair comparison between MaNGA and IllustrisTNG, some small differences remain which could lead to minor discrepancies between the resulting BTFRs. For example, the samples assume a slightly different cosmology and IMF. However, given that these factors impact mass estimates, but not 21cm linewidths, any offset should only be present in the BTFR zero points, not the slopes.  Despite the expected differences, both slopes and intercepts agree within $1\sigma$. We expect this because systematic shifts due to assumed IMF and cosmology are smaller than typical errors on determinations of stellar mass. 

In the following, we demonstrate the impact that using mock observations has on the calibration of the BTFR for simulated data. We also present a comparison of our results with other observational studies.

\subsection{The necessity of mock observations: the IllustrisTNG BTFR with rotation velocity directly from IllusrisTNG}
\label{sec:need_mocks}

\begin{figure*}
\centering
\includegraphics[angle=0,scale=0.3]{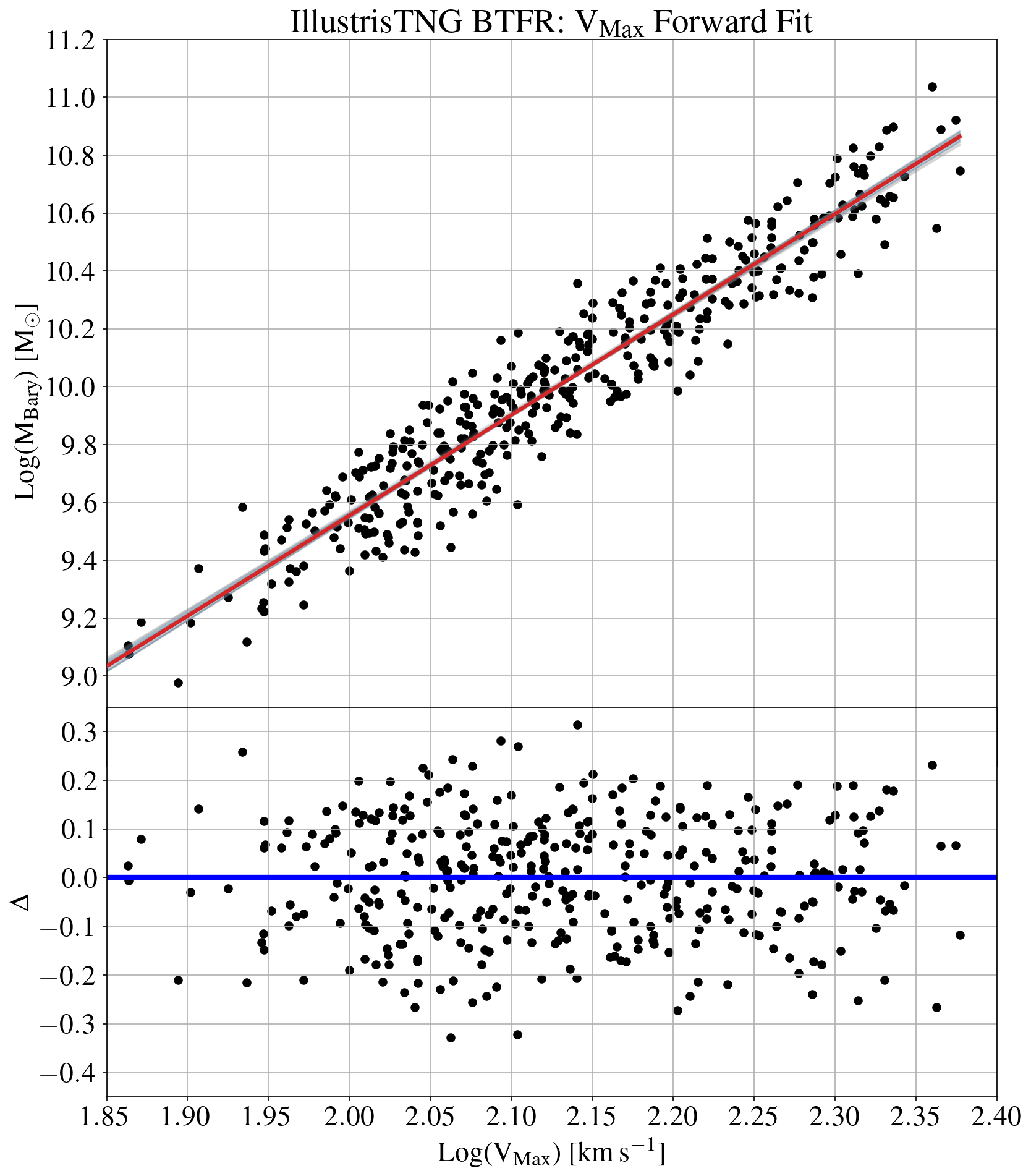}
\includegraphics[angle=0,scale=0.3]{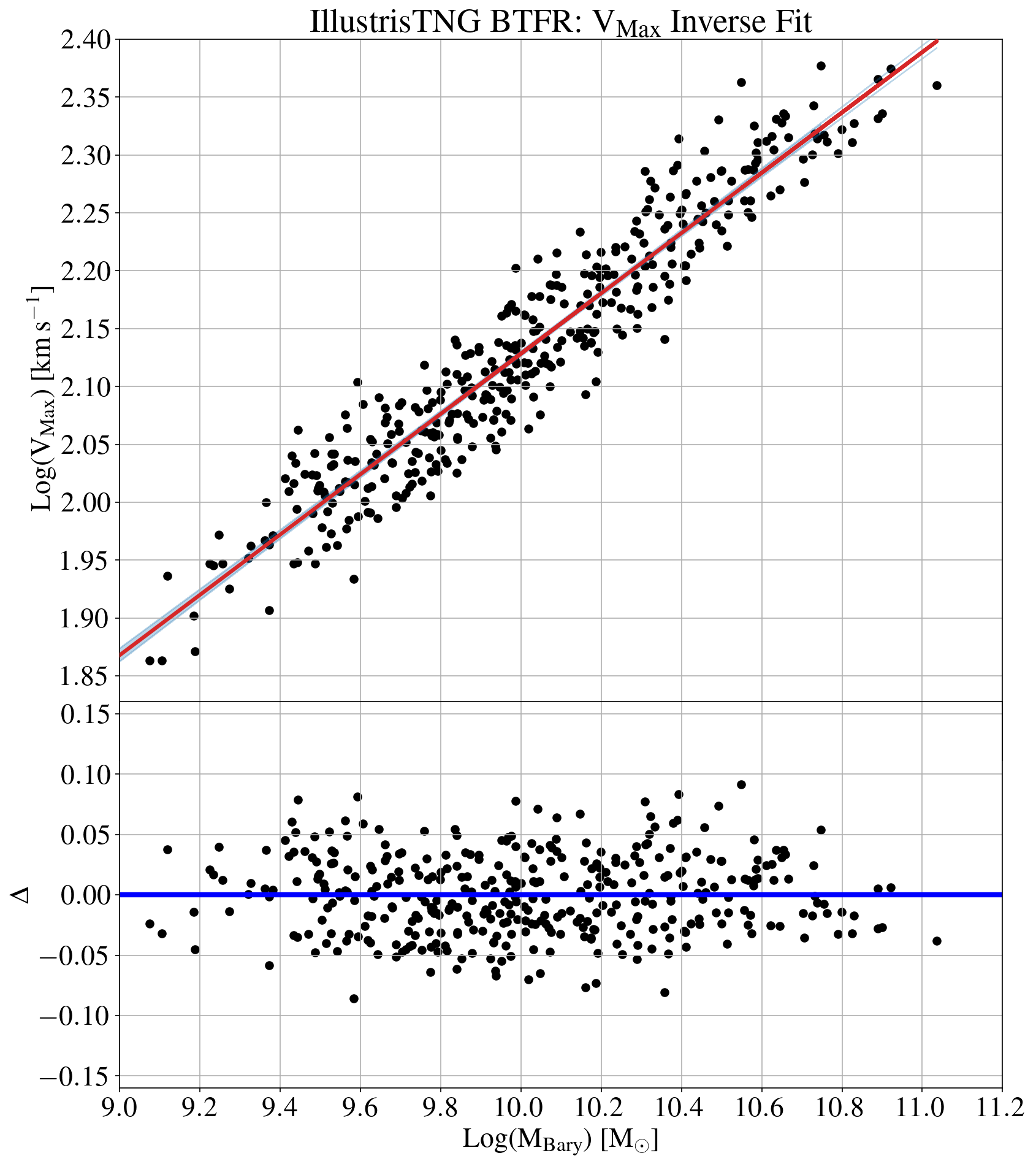}
\includegraphics[angle=0,scale=0.3]{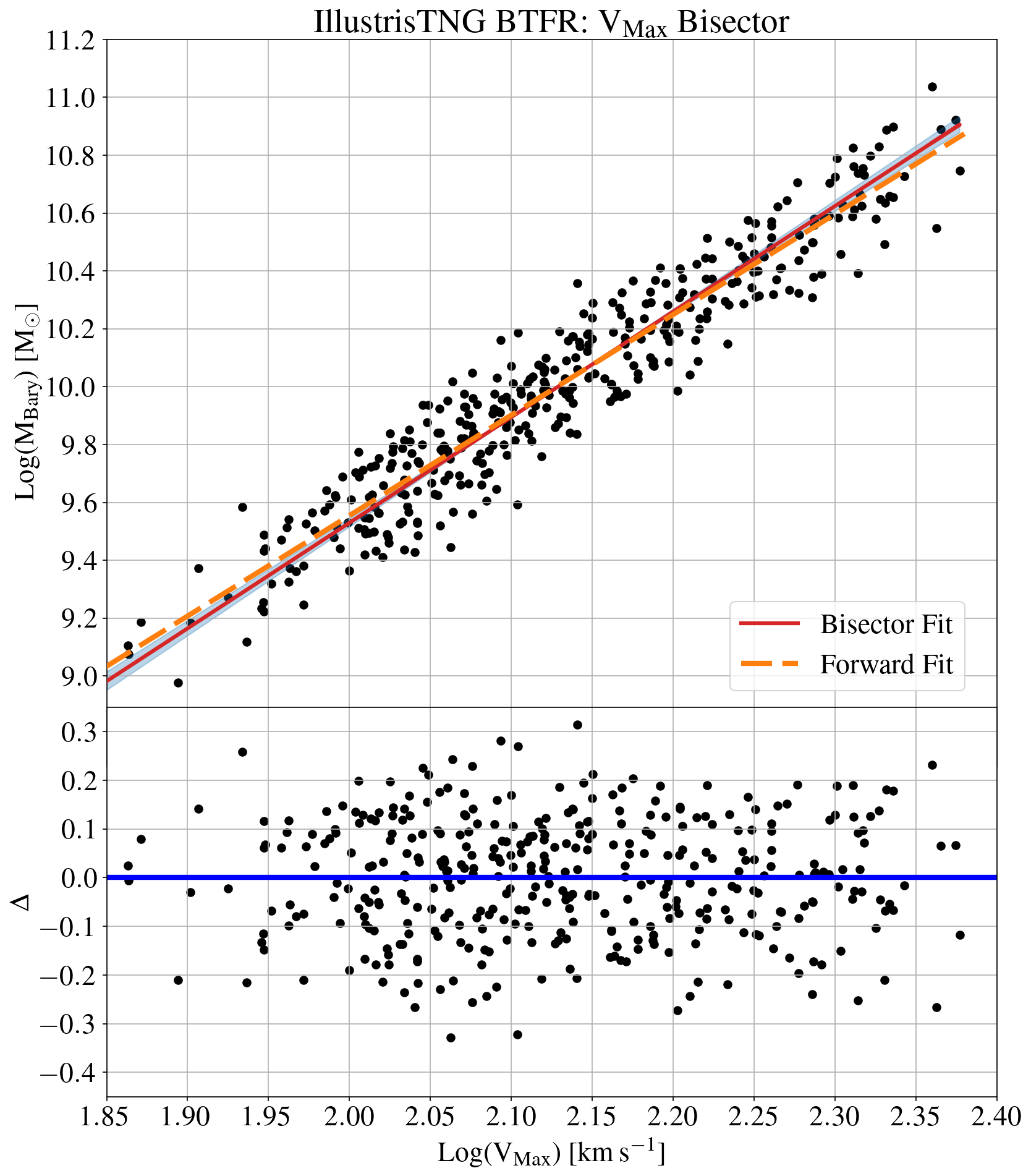}

\caption{The IllustrisTNG BTFR with the IllustrisTNG rotation velocities directly rather than using mock observations. The top left shows rotation velocity on the horizontal axis and baryonic mass on the vertical axis. The top right shows baryonic mass on the horizontal axis and rotation velocity on the vertical axis. The bottom shows the bisector fit. We use the same color criteria as in Figures \ref{figure:MaNGA_BTFR} and \ref{figure:TNG_BTFR}. The best fit lines, clockwise from left, have slope 3.47 $\pm$ 0.07 with a y-intercept of 2.61 $\pm$ 0.14, slope of 3.84 $\pm$ 0.08 with a y-intercept of 1.83 $\pm$ 0.17, and 3.65 $\pm$ 0.11 with a y-intercept of 2.25 $\pm$ 0.12. We note that the mass and rotation velocity measurements have zero uncertainty because they are taken directly from IllustrisTNG and are known perfectly since they are simulated. This BTFR does not agree with MaNGA, demonstrating the need for the mock observations to create a fair and consistent comparison between observation and theory.}
\label{figure:TNG_vmax_BTFR}
\end{figure*}

\begin{figure}
\centering
\includegraphics[angle=0,scale=0.3]{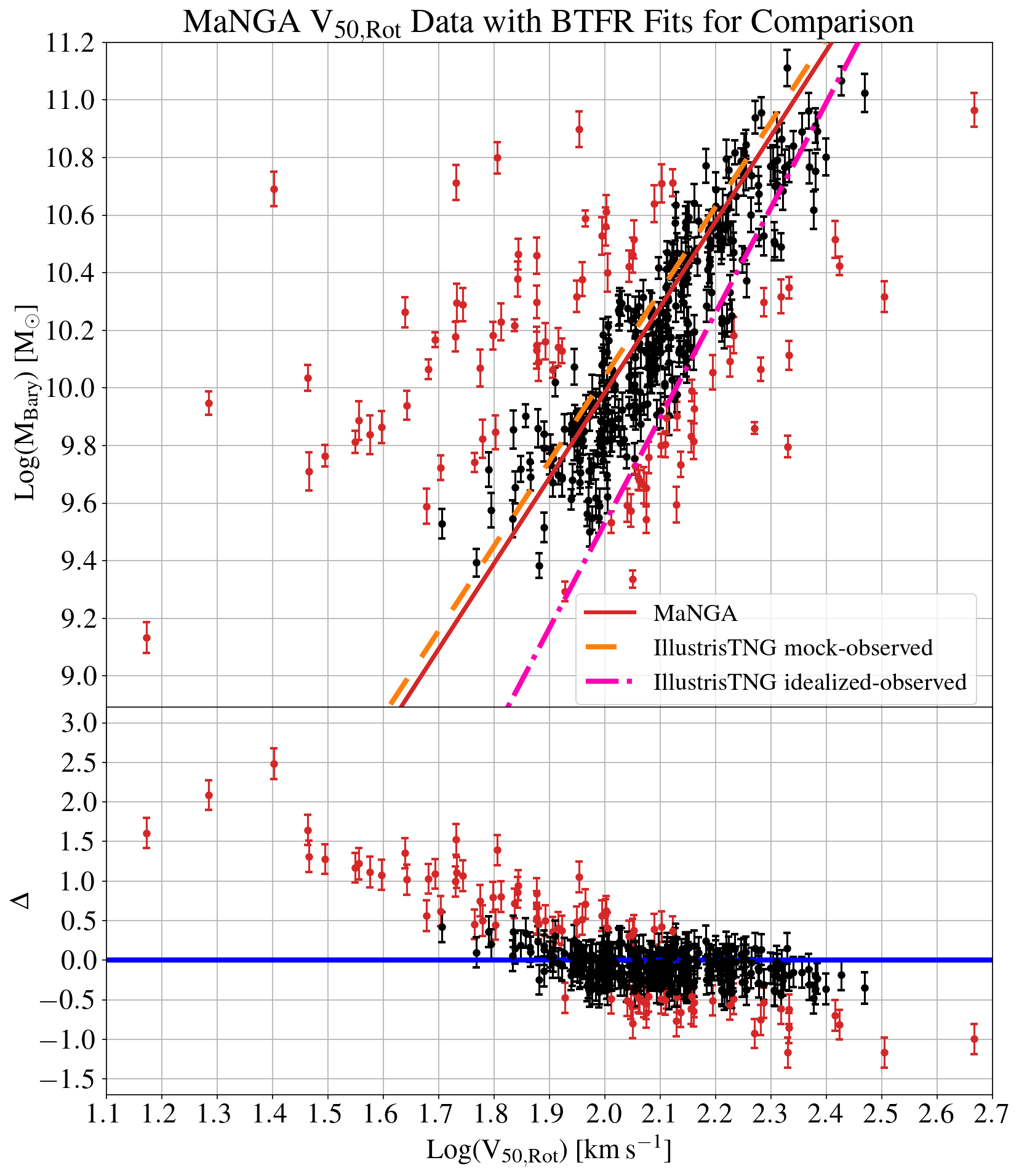}
\caption{The MaNGA, IllustrisTNG mock-observed, and IllustrisTNG direct BTFR fits on the MaNGA data. The solid red line is the MaNGA BTFR fit. the dashed orange line is the IllustrisTNG mock-observed BTFR fit. The dash-dotted pink line is the BTFR directly from IllustrisTNG. This shows that the BTFR measured directly from IllustrisTNG does not agree the one from MaNGA but the mock observed IllustrisTNG BTFR does. This demonstrates the need for the mock observations to create a fair and consistent comparison between observation and theory.}
\label{figure:BTFR_fit_comparison}
\end{figure}

A major emphasis of this work is the use of mock HI observations of IllustrisTNG galaxies to ensure a fair comparison with a real observational sample. We now demonstrate the impact and necessity for mock observations of simulated data when making comparisons with observational data. Specifically, we do not use the mock GBT observations (see Section \ref{section:BTFRinTNG}). Instead, we create a BTFR using the rotational velocities extracted directly from IllustrisTNG and HI masses created through idealized observations that do not apply the GBT observational uncertainties. We employ the same bisector method used to derive the MaNGA and IllustisTNG BTFR's with the mock observations. We show the rotation velocity vs. baryonic mass and baryonic mass vs. rotation velocity fits for the IllustrisTNG galaxies, as well as the corresponding bisector fit, in Figure \ref{figure:TNG_vmax_BTFR} using the same color criteria as in Figures \ref{figure:MaNGA_BTFR} and \ref{figure:TNG_BTFR}. The rotation velocity vs. baryonic mass fit has a slope of 3.47 $\pm$ 0.07 with a y-intercept of 2.61 $\pm$ 0.14 $\log_{10}M_{\odot}$. The baryonic mass vs. rotation velocity fit has a slope of 3.84 $\pm$ 0.08 with a y-intercept of 1.82 $\pm$ 0.17 $\log_{10}M_{\odot}$. The corresponding bisector fit has a slope of 3.65 $\pm$ 0.11 with a y-intercept of 2.25 $\pm$ 0.12 $\log_{10}M_{\odot}$ These results are also provided in Table~\ref{tab:BTFRwithAndWithoutTurbulence}. We note that the  measurements from IllustrisTNG are known perfectly since they are simulated and consequently have zero uncertainty. Although this BTFR has less scatter and consequently smaller uncertainties, it notably does not agree with the MaNGA BTFR, showing a significantly steeper bisector slope. To emphasize this disagreement, we plot the MaNGA, IllustrisTNG mock-observed, and IllustrisTNG direct BTFR fits on the MaNGA data. This is shown in Figure \ref{figure:BTFR_fit_comparison} using the same color criteria as in Figures \ref{figure:MaNGA_BTFR} and \ref{figure:TNG_BTFR}. There is no expectation that the BTFR based on theoretical maximum velocity should match that based on observed line width \citep[e.g. see][]{Lelli2019}. We confirm here that they do not match, and therefore mock observations of simulations are important for matching with real observed relations.

\subsection{A Comparison to the Literature}
\label{sec:comparison}
Due to the various definitions of rotation velocity, we can only accurately compare our results to other studies that also use linewidths (see Section \ref{sec:Intro} for details on these different definitions). The slope of the BTFR derived using linewidths in previous studies vary between 3 and 4 \citep{Noordermeer2007, Avila-Reese2008, Gurovich2010, Zaritsky2014, Bradford2016, Brook2016, Papastergis2016, Lelli2019}. The MaNGA and IllustrisTNG BTFR slopes are both low in comparison with the literature, but still consistent within uncertainties. The comparison with these authors is shown in Table~\ref{tab:BTFRcomparison}. 

There are additional differences between studies that could affect our comparisons, such as linewidth definition, turbulent velocity corrections, and sample definition. In this section, we test how these choices can affect our resulting BTFR. Our alternative BTFR fits are given in Table~\ref{tab:BTFRwithAndWithoutTurbulence}. 

Many studies represent velocity using $W_{20}$, rather than $W_{50}$. To test the impact of this choice, we calculate the best-fit BTFR using $V_{20}$ instead of $V_{50}$. We find using $V_{20}$ has negligible impact on the the MaNGA BTFR. The IllustrisTNG BTFR slope increases mildly by 0.15, which is within the uncertainties of the original fit.

While it is common for measured 21cm linewidths to be corrected for instrumental broadening \citep{Springob2005}, many previous studies do not always apply a correction for turbulent motions. We explore whether our shallow slope compared with the literature can be attributed to the unequal influence of turbulence across all parameter space. Our corrected linewidths (see Eq. \ref{eq:Wrot}) include a constant factor, $\Delta t$, which is the turbulent velocity correction. As this factor is constant, it has a larger effect on less massive galaxies which host smaller linewidths, preferentially decreasing the rotation velocity of smaller galaxies and possibly biasing the best fit to shallower slopes.  To test its impact, we reran our fits without the turbulence correction, but found it has only a marginal effect on the derived BTFR and consequently does not account for our shallow slope, as shown in Table~\ref{tab:BTFRwithAndWithoutTurbulence}.

Similarly, while previous studies are often explicitly limited to disk galaxies, we have made no morphology cut, although our sample will be biased towards disk galaxies by design of requiring HI detections. Nonetheless, there is a possibility that including early-type galaxies, which have less rotational support, could be responsible for our shallower BTFR slope. To test this possibility, we employ the $\mu_{\Delta}$ parameter of \ {Kannappan2013} to identify and reject spheroid dominated galaxies from our fits. The $\mu_{\Delta}$ quantity is a stellar mass surface density contrast between the regions within and external to $r_{50}$, similar to a concentration index, but providing a cleaner separation between early- and late-type galaxies \citep{Moffett2015}. We exclude early-type galaxies by rejecting all galaxies with $\mu_{\Delta} > 9.5$. We note that this cut is based on the sample in  \citet{Kannappan2013}, and we assume it is valid for our sample as well. Limiting our sample in this way has negligible impact on the resulting slopes for our MaNGA sample.  The slopes derived from the IllustrisTNG sample are somewhat higher by $\sim0.2$. However, the MaNGA and IllustrisTNG samples still provide results that agree with each other within the uncertainties.

\begin{table*}
 \centering
 \begin{tabular}{|l|c|c|c|c|c|}
 \hline
 Type of BTFR Fit & $\Delta t$ Correction? &  Sample Size & Number of Outliers & Slope$^{\dagger}$ & Zero Point$^{\dagger}$\\
  & &  & &  & $\log{M_{\rm Bary}/M_{\odot}}$ \\
 \hline
 \hline
 \multicolumn{6}{c}{Main Comparison ($V_{\rm 50,Rot}$ and $M_{\rm Bary}$)} \\
\hline
 MaNGA: Forward  & Yes  & 377 & 88 &  2.59 $\pm$ 0.09 & 4.76 $\pm$ 0.19 \\
 MaNGA: Inverse  & Yes  & 377 & 9  &  3.44 $\pm$ 0.16 & 2.95 $\pm$ 0.33\\
 MaNGA: Bisector & Yes  & 377 & 88 &  2.97 $\pm$ 0.18 & 4.04 $\pm$ 0.41 \\
 \hline
 IllustrisTNG: Forward  & Yes  & 377 & 78 &  2.50 $\pm$ 0.13 & 4.94 $\pm$ 0.28 \\
 IllustrisTNG: Inverse  & Yes  & 377 & 2  &  3.56 $\pm$ 0.28 & 2.78 $\pm$ 0.57 \\
 IllustrisTNG: Bisector & Yes  & 377 & 78 &  2.94 $\pm$ 0.23 & 4.15 $\pm$ 0.44 \\
  \hline
 \hline
\multicolumn{6}{c}{Without mock observations ($V_{\rm max}$ and $M_{\rm Bary}$)} \\
 \hline
IllustrisTNG: Forward & N/A & 377 & 0  & 3.47 $\pm$ 0.07 & 2.61 $\pm$ 0.14 \\
IllustrisTNG: Inverse & N/A  & 377 & 0  & 3.84 $\pm$ 0.08 & 1.82 $\pm$ 0.17 \\
IllustrisTNG: Bisector & N/A &  377 & 0  & 3.65 $\pm$ 0.11 & 2.25 $\pm$ 0.12 \\ 
\hline
\hline 
\multicolumn{6}{c}{Testing HI width choice ($V_{\rm 20,Rot}$ and $M_{\rm Bary}$)} \\
\hline
MaNGA: Forward   & Yes &  377 & 40 & 2.51 $\pm$ 0.17 & 4.78 $\pm$ 0.37 \\
 MaNGA: Inverse  & Yes & 377 & 5  &  3.73 $\pm$ 0.18 & 2.17 $\pm$ 0.40\\
MaNGA: Bisector & Yes & 377 & 40 &  3.01 $\pm$ 0.17 & 3.73 $\pm$ 0.36 \\
 \hline
IllustrisTNG: Forward & Yes  & 377 & 39 &  2.58 $\pm$ 0.10 & 4.85 $\pm$ 0.22 \\
IllustrisTNG: Inverse & Yes &  377 & 11 &  3.82 $\pm$ 0.26 & 2.22 $\pm$ 0.55 \\
IllustrisTNG: Bisector & Yes  & 377 & 39 & 3.09
 $\pm$ 0.26 & 3.68 $\pm$ 0.51 \\
 \hline
 \hline
\multicolumn{6}{c}{Testing turbulence corrections ($V_{\rm 50,Rot}$ and $M_{\rm Bary}$)}\\
\hline
 MaNGA: Forward  & No &  377 & 87 &  2.75 $\pm$ 0.10 & 4.39 $\pm$ 0.21 \\
 MaNGA: Inverse  & No &  377 & 7 &  3.57 $\pm$ 0.17 & 2.64 $\pm$ 0.36 \\
 MaNGA: Bisector & No &  377 & 87 &  3.11 $\pm$ 0.23 & 3.70 $\pm$ 0.40 \\
 \hline
 IllustrisTNG: Forward  & No  & 377 & 70  & 2.48 $\pm$ 0.20 & 4.96 $\pm$ 0.42 \\
 IllustrisTNG: Inverse  & No &  377 & 2  & 3.74 $\pm$ 0.31 & 2.36 $\pm$ 0.63 \\
 IllustrisTNG: Bisector & No &  377 & 70  & 2.97 $\pm$ 0.22 & 4.12 $\pm$ 0.60 \\
 \hline
 \hline
 \multicolumn{6}{c}{Morphology cut, $\mu_\Delta<9.5$ to select disks ($V_{\rm 50,Rot}$ and $M_{\rm Bary}$)} \\
 \hline
MaNGA: Forward & Yes  & 368 & 86 & 2.59 $\pm$ 0.09 & 4.77 $\pm$ 0.19 \\
MaNGA: Inverse & Yes &  368 & 9 & 3.47 $\pm$ 0.16 & 2.91 $\pm$ 0.34\\
 MaNGA: Bisector & Yes & 368 & 86 & 2.97 $\pm$ 0.19 & 4.05 $\pm$ 0.41 \\
 \hline
IllustrisTNG: Forward & Yes  & 330 & 72 & 2.71 $\pm$ 0.12 & 4.47 $\pm$ 0.25 \\
IllustrisTNG: Inverse & Yes &  330 & 0 &  3.68 $\pm$ 0.30 & 2.51 $\pm$ 0.63 \\
IllustrisTNG: Bisector &  Yes & 330 & 72  & 3.13 $\pm$ 0.22 & 3.73 $\pm$ 0.44 \\
\hline
 \hline 
 \end{tabular}
 \caption{Our various BTFR fits, for V$_{\rm 50}$ and V$_{\rm 20}$, testing the impact of HI width choice, use of mock observations, a morphological $\mu_\Delta$ cut and turbulence corrections.$^{\dagger}$The inverse fits are initially calculated as $V_{\rm Rot}=m'M_{\rm bary}+b'$, but we present the slope and zero points after converting the results in the ``forward fit" form of $M_{\rm bary}=mV_{\rm rot}+b$, where $m=1/m'$ and $b=-b'/m'$.  For all of the permutations, the MaNGA and IllustrisTNG fits agree within uncertainties.}
 \label{tab:BTFRwithAndWithoutTurbulence}
\end{table*}

\begin{table}
\begin{tabular}{cccc}
\hline
Line   & $\Delta t$ & Slope             & Reference    \\ 
width & correction? &  \\
\hline       
$W_{\rm F50}$   & Yes                    & 2.97 $\pm$ 0.18   & This work (MaNGA data/bisector)          \\
$W_{\rm F50}$   & Yes                    & 2.97 $\pm$ 0.22   & This work (Illustris TNG/bisector)       \\
$W_{\rm F20}$   & Yes                    & 3.01 $\pm$ 0.17   & This work (MaNGA data/bisector)         \\
$W_{\rm F20}$   & Yes                   & 3.09 $\pm$ 0.26   & This work (Illustris TNG/bisector)       \\
$W_{\rm 20}$    & Yes                    & $ 3.27 \pm 0.13 $ & \citet{Avila-Reese2008} \\
$W_{\rm 50}$    & Yes                    & 3.5 $\pm$ 0.2     & \citet{Zaritsky2014}    \\
$W_{\rm M50}/2$ & Yes                    & 3.62 $\pm$ 0.09   & \citet{Lelli2019}       \\
\hline
$W_{\rm 50}/2$  & No                     & 2.70              & \citet{Brook2016}       \\
$W_{\rm 50}$    & No?                    & 2.80 $\pm$ 0.14 & \citet{Ponomareva2018} \\ 
$W_{\rm F50}$   & No                     & 2.97 $\pm$ 0.29   & This work (IllustrisTNG/bisector)       \\
$W_{\rm 20}$    & No                     & $ 3.04 \pm 0.08 $ & \citet{Noordermeer2007} \\
$W_{\rm F50}$   & No                     & 3.11 $\pm$ 0.19   & This work (MaNGA data/bisector)          \\
$W_{\rm 20}/2$  & No                     & 3.13              & \citet{Brook2016}       \\
$W_{\rm 20}$    & No                     & $ 3.20 \pm 0.10 $ & \citet{Gurovich2010}    \\
$W_{\rm 20}$    & No                     & 3.24 $\pm$ 0.05   & \citet{Bradford2016}    \\
$W_{\rm 50}/2$  & No                     & 3.75 $\pm$ 0.11   & \citet{Papastergis2016} \\
$W_{\rm P20}/2$ & No                     & 3.85 $\pm$ 0.09   & \citet{Lelli2019}      \\
 \hline
 \hline
\end{tabular}
 \caption{Our various BTFR fits showing that our slopes are shallow compared with some from the literature also using linewidths. $W_{20}$ is the linewidth at 20\% of its peak and $W_{50}$ is the linewidth at 50\% of its peak.}
 \label{tab:BTFRcomparison}
\end{table}

\section{Conclusions and Future Directions}
\label{sec:conclusions}

We compare the observed BTFR using SDSS-IV MaNGA and HI-MaNGA observations to a simulated one from IllustrisTNG. We perform mock 21 cm observations on galaxies from IllustrisTNG, ensuring a fair comparison between the observed galaxies with observational limitations, biases, and uncertainties and the simulation without those constraints, as well as to minimize disagreements caused by different measurement methods. We report a MaNGA BTFR slope of 2.97 $\pm$ 0.18 with a y-intercept of 4.04 $\pm$ 0.41 and an IllustrisTNG BTFR slope of 2.94 $\pm$ 0.23 with a y-intercept of 4.15 $\pm$ 0.44. Thus, MaNGA and IllustrisTNG produce BTFRs that agree within uncertainties, suggesting IllustrisTNG has created a galaxy population that accurately represents that of the real universe when judged by BTFR. 

To demonstrate the importance of mock observations on simulated data, we also create a BTFR using rotation velocity directly from IllustrisTNG and demonstrate that it is necessary to create the BTFR in a fair and consistent way between observation and theory when comparing the two.  This is noteworthy because simulations have historically struggled to reproduce observed galaxy scaling relations such as the BTFR \citep{Somerville2015}.

IllustrisTNG is used to simulate a wide range of cosmological and extra-galactic phenomena, including thermodynamic structure in galaxy clusters \citep{Barnes2018}, the large-scale distribution of highly ionized metals \citep{Artale2021}, 
dark matter halo formation \citep{Hearin2021}, the effect of galaxy mergers on galaxy evolution \citep{Hani2020}, the dark matter and the star formation in galaxies \citep{Lovell2018, Tacchella2019}, the properties and distribution of bars in spiral galaxies \citep{Zhao2020}, supermassive black hole formation and feedback \citep{Weinberger2018}, and even fast radio bursts \citep{Zhang2021} and binary neutron star mergers \citep{Rose2021}. This demonstration that IllustrisTNG accurately recreates the observed BTFR gives us more confidence that those phenomena simulated with IllustrisTNG actually reflect reality. 

There are a number of avenues for expanding upon our analysis. This work primarily focused on the treatment of single dish 21cm spectra, and future work may benefit from incorporating and comparing results from alternative linewidth estimation algorithms, as has been done other studies \citep{Trachternach2009, McGaugh2012, Desmond2015, Bradford2016, Brook2016, Lelli2019, Glowacki2020}. Estimating linewidths using measured rotation curves from MaNGA IFU spectroscopy, and mock rotation curves from IllustrisTNG galaxies is a natural next step. Lastly, future analysis may benefit from a careful homogenization in the methods used to calculate stellar masses in MaNGA and IllustrisTNG, in which mock spectrophotometry of IllustrisTNG galaxies are generated and passed through the same programs used to estimate the stellar masses in the observational data. Such analyses should be possible thanks to routines that can postprocess simulated data into observed spectra \citep[][]{Torrey15}. More HI-MaNGA data is coming soon, which will give us a larger data set to perform these analyses.

\section*{Acknowledgements}

This research was conducted through the Haverford College Koshland Integrated Natural Sciences Center (KINSC). We would like to acknowledge the ALFALFA teams. We would additionally like to thank Dr. Coleman Krawczyk for his Bayesian MCMC linear fitting algorithm and helping us adapt it. This work also makes use of the MaNGA-Pipe3D data products. We thank the IA-UNAM MaNGA team for creating the Pipe3D catalogue, and the Conacyt Project CB-285080 for supporting them.

Funding for the Sloan Digital Sky Survey IV has been provided by the Alfred P. Sloan Foundation, the U.S. Department of Energy Office of Science, and the Participating Institutions.

SDSS-IV acknowledges support and resources from the Center for High Performance Computing at the University of Utah. The SDSS website is www.sdss.org.

SDSS-IV is managed by the Astrophysical Research Consortium for the Participating Institutions of the SDSS Collaboration including the Brazilian Participation Group, the Carnegie Institution for Science, Carnegie Mellon University, Center for Astrophysics | Harvard \& Smithsonian, the Chilean Participation Group, the French Participation Group, Instituto de Astrof\'isica de Canarias, The Johns Hopkins University, Kavli Institute for the Physics and Mathematics of the Universe (IPMU) / University of Tokyo, the Korean Participation Group, Lawrence Berkeley National Laboratory, Leibniz Institut f\"ur Astrophysik Potsdam (AIP), Max-Planck-Institut f\"ur Astronomie (MPIA Heidelberg), Max-Planck-Institut f\"ur Astrophysik (MPA Garching), Max-Planck-Institut f\"ur Extraterrestrische Physik (MPE), National Astronomical Observatories of China, New Mexico State University, New York University, University of Notre Dame, Observat\'ario Nacional / MCTI, The Ohio State University, Pennsylvania State University, Shanghai Astronomical Observatory, United Kingdom Participation Group, Universidad Nacional Aut\'onoma de M\'exico, University of Arizona, University of Colorado Boulder, University of Oxford, University of Portsmouth, University of Utah, University of Virginia, University of Washington, University of Wisconsin, Vanderbilt University, and Yale University. 

This work also makes use of the NumPy \citep{NumPy2020}, Matplotlib \citep{Matplotlib} , Astropy \citep{astropy2013, astropy2018}, Pymc3 \citep{Salvatier2015} and Theano \citep{Theano2016} Python models.

The Green Bank Observatory is a facility of the National Science Foundation operated under cooperative agreement by Associated Universities, Inc.

\section*{Data Availability}

The HI-MaNGA catalog used in this study can be found at  \url{https://greenbankobservatory.org/science/gbt-surveys/hi-manga/}. The MaNGA MPL-10 data products can be generated by the public using the raw data (available at \url{https://www.sdss.org/dr16/manga/manga-data/data-access/)} with DRP v3.0.1 and Pipe3D v3.0.1. IllustrisTNG files are available from \url{https://www.tng-project.org/data/} and mock HI data cubes can be recreated using the Martini package (\url{https://github.com/kyleaoman/martini}).


\bibliographystyle{mnras}
\bibliography{references} 

\bsp	
\label{lastpage}
\end{document}